\documentclass[aps,pra,twocolumn,amsmath,amssymb]{revtex4-2}
\usepackage[siunitx]{circuitikz}
\usepackage{graphicx}
\usepackage{subcaption}
\usepackage{bm}
\usepackage{physics}
\usepackage{float}
\usepackage{comment}
\usepackage[normalem]{ulem}
\begin{document}

\title{Left-Handed Transmission Lines for the Dynamical Casimir Effect}
\author{Valentin E. Salari}
\email{ve.salari@df.uba.ar}
\author{Fernando C. Lombardo}
\email{lombardo@df.uba.ar}
\author{Paula I. Villar}
\email{paula@df.uba.ar}
\affiliation{Departamento de Física Juan José Giambiagi and IFIBA UBA-CONICET, Facultad de Ciencias Exactas y Naturales, Ciudad Universitaria, Pabellón I, 1428 Buenos Aires, Argentina}
\date{\today}

\begin{abstract}
Left-handed (LH) superconducting transmission lines, built with series capacitors and shunt inductors instead of the series inductors and shunt capacitors of a conventional right-handed (RH) line, have a spectrum with a nonzero lower band edge and modes that become sparse at high frequencies. 
We study the dynamical Casimir effect in a LH line terminated by a flux-modulated SQUID and compare it with an RH line built from the same elements and with a long coplanar waveguide terminated by the same SQUID. 
We derive the quantum Hamiltonian and show that the coupling between two modes factorizes into their amplitudes at the SQUID.
 Together, the LH spectrum and the mode dependence of this coupling determine which modes are populated and how fast. 
With equal elements, the LH coupling grows with frequency and exceeds the RH one above a crossover pump frequency set by the cell inductance and capacitance, and only the LH line has resonances above the RH band edge. Pumping two modes at once lets a weakly coupled mode borrow coupling from a strongly coupled one, while crossings of resonance lines populate three or four modes without increasing the growth rate. Compared with the coplanar waveguide, the LH line has a similar coupling but keeps the created particles in the target mode, whereas the nearly equidistant waveguide spreads them over many pairs and needs more pump energy.
\end{abstract}

\maketitle

\section{Introduction}
\label{sec:intro}

Quantum vacuum fluctuations are one of the most striking manifestations of quantum field theory, giving rise to macroscopic phenomena such as the static Casimir force between conducting boundaries~\cite{Casimir1948}. When boundary conditions are modulated rapidly in time, the quantum vacuum becomes unstable and converts virtual excitations into real, measurable particles—a phenomenon known as the dynamical Casimir effect (DCE)~\cite{Moore1970, Law1994, DodonovKlimov1996, Crocce2001, DalvitMaiaNetoMazzitelli2011, Nation2012, Dodonov2020}. For decades, observing the DCE via the mechanical motion of mirrors remained out of experimental reach, as generating a detectable photon flux requires boundary velocities approaching a significant fraction of the speed of light~\cite{Nation2012, LombardoVillar2025}.

Circuit quantum electrodynamics (cQED)~\cite{YurkeDenker1984, Blais2004, Wallraff2004, VoolDevoret2017, Blais2021} provided a breakthrough platform to overcome this limitation by replacing mechanically moving mirrors with tunable electromagnetic boundary conditions. By terminating a superconducting transmission line with a superconducting quantum interference device (SQUID) and modulating the external magnetic flux threading its loop at gigahertz frequencies, one can rapidly vary the effective Josephson inductance at the boundary~\cite{Johansson2009, JohanssonCasimir2010, Wustmann2013, Lombardo2016, Lombardo2018}. This modulates the effective electrical length of the waveguide at relativistic speeds, enabling the first experimental observation of the DCE in open transmission lines~\cite{Wilson2011} as well as parametric photon creation in closed superconducting resonators~\cite{Wilson2010,Lahteenmaki2013, Svensson2018}. Beyond fundamental tests of quantum vacuum radiation, SQUID-terminated waveguides and Josephson metamaterials have become central building blocks for quantum-limited parametric amplification, two-mode squeezing, and continuous-variable entanglement generation~\cite{CastellanosBeltran2008, Macklin2015, Schneider2020}, and the DCE has been proposed as a resource for quantum gates and quantum thermal machines in cQED~\cite{DelGrosso2022Otto, DelGrosso2025Gate}. More generally, superconducting transmission lines have become versatile quantum simulators of $(1+1)$-dimensional quantum fields, able to reproduce particle creation and quantum-optical processes driven by the modulation of circuit elements~\cite{Nation2012, FerreriWilhemCreation, FerreriWilhemSimulation, Urtubey2026}.

In standard cQED architectures, the waveguide is fabricated as a continuous, right-handed (RH) coplanar transmission line. While convenient to fabricate, a finite continuous line exhibits a linear dispersion relation at low frequencies, resulting in a spectrum of normal modes that is practically equidistant~\cite{JohanssonCasimir2010, Wustmann2013, Lombardo2016}. Consequently, driving the boundary at twice the fundamental mode frequency  excites higher-order harmonics and inter-mode conversion processes~\cite{DodonovKlimov1996, Crocce2001, Lombardo2016, Lombardo2018}, complicating the selective control of isolated quantum states, and, as we show in Sec.~\ref{sec:LvsRperfo}, increasing the pump energy needed to populate a given mode. Moving to a discrete, lumped-element transmission line composed of a finite number of unit cells $N$ naturally breaks this harmonic equidistance due to lattice dispersion~\cite{EggerWilhem2013, FerreriWilhemCreation}, while simultaneously introducing a rich modal structure governed by the discrete boundary conditions at the SQUID.

Furthermore, discrete circuit architectures allow for the topological inversion of the reactive elements in each unit cell~\cite{Eleftheriades2002, LaiItohCaloz2004, Caloz2004, CalozItohBook, Kozyrev2008}. Whereas a standard right-handed line consists of series inductors and shunt capacitors to ground, a left-handed (LH) metamaterial transmission line places capacitors in series and inductors shunting each node to ground. This role reversal leads to an anomalous dispersion relation where phase and group velocities point in opposite directions, supporting backward waves~\cite{Veselago1968, Smith2000, Pendry2000, Houck2003}. Crucially for parametric processes, whereas the spectrum of a discrete RH line is bounded from above by a cutoff frequency and becomes dense at high frequencies, a purely LH line is bounded from below by an infrared cutoff and extends toward high frequencies at long wavelengths, where the modes become strongly non-equidistant~\cite{EggerWilhem2013, FerreriWilhemCreation, FerreriWilhemSimulation}. Superconducting LH lines have been fabricated and coupled to transmon qubits, displaying the predicted dense mode spectrum above the infrared cutoff~\cite{Wang2019, Indrajeet2020, McBroomCarroll2024}, and have been proposed as multimode buses and quantum simulators in cQED~\cite{EggerWilhem2013, Messinger2019}. Particle creation in LH lines has been studied in Ref. \cite{FerreriWilhemCreation}, with SQUIDs in every cell and a sudden change of their Josephson energy. Here we consider a single SQUID at the end of the line, modulated periodically, as in DCE experiments \cite{Wilson2010,Wilson2011} and theoretical studies \cite{Wustmann2013,Lombardo2016}.

In this work, we investigate and compare the dynamical Casimir effect in discrete left-handed (LHST) and right-handed (RHST) transmission lines terminated by a flux-modulated SQUID, operating with a finite number of unit cells $N$ without taking the continuum limit ($N \to \infty$).  We use the same elements in both lines, so that the comparison isolates the effect of exchanging the series and shunt elements. Starting from the exact circuit topology, we derive the canonical quantization and the full parametric interaction Hamiltonian for both architectures, showing that the inter-mode coupling constants factorize in terms of the spatial mode amplitudes at the SQUID node. We demonstrate that the dual dispersion relations and boundary conditions lead to opposite frequency dependencies in the parametric couplings: while the single-mode coupling in the RHST line remains bounded and nearly flat across its band, the LHST coupling grows steeply with frequency. As a result, although the LHST line attenuates particle creation near its lower band edge, it strongly enhances particle creation at high frequencies and provides access to spectral regimes beyond the RHST cutoff. Moreover, we show that the low-frequency attenuation of the LHST line can be overcome via two-mode resonance pumping, where a weakly coupled low-frequency target mode borrows coupling strength from a strongly coupled high-frequency partner. We then explore multimode resonances arising from non-uniform shifts of the transcendental eigenfrequencies—leading to three- and four-mode crossings—and show analytically and numerically that in the dense spectral region near the Brillouin zone edge, the coupling strength decays faster with $N$ than the frequency gap ($\mathcal{O}(N^{-3})$ versus $\mathcal{O}(N^{-2})$), preventing uncontrolled multimode instabilities in long arrays. Finally, we compare both lines as a function of the pump frequency, which is the parameter that can be tuned in an experiment once the circuit is fabricated, and show that they exchange roles at a crossover frequency $\Omega_c \simeq 2/\sqrt{LC}$ set only by the cell elements: below it the RHST line creates particles faster, above it the LHST line does, and above the RHST band edge only the LHST line has resonances. We also compare the LHST line with a long coplanar waveguide terminated by the same SQUID: with similar couplings, the nearly equidistant waveguide drives many pairs together with the target and needs more pump energy, whereas the LHST line keeps the created particles in the target mode.

The paper is organized as follows. In Sec.~\ref{sec:model}, we present the circuit models for the LHST and RHST lines, derive their normal modes, transcendental boundary equations and quantized Hamiltonian, and compare their spectra and coupling distributions. Section~\ref{sec:singleresonance} addresses particle creation under single-mode (Sec.~\ref{1res}) and two-mode (Sec.~\ref{2res}) resonant pumping, and Sec.~\ref{sec:multipleresonance} studies crossings of resonance lines as a function of the circuit inductance and the resulting three- and four-mode dynamics. In Sec.~\ref{sec:densespectrum}, we analyze the asymptotic scaling of couplings and frequency spacings in dense spectral regions. In Sec.~\ref{sec:LvsRperfo}, we compare both lines as a function of the pump frequency and contrast the LHST line with a long coplanar waveguide terminated by the same SQUID. Finally, we summarize our conclusions in Sec.~\ref{sec:conclusion}.

\section{Left- and Right- Handed SQUID Terminated Transmission line}
\label{sec:model}

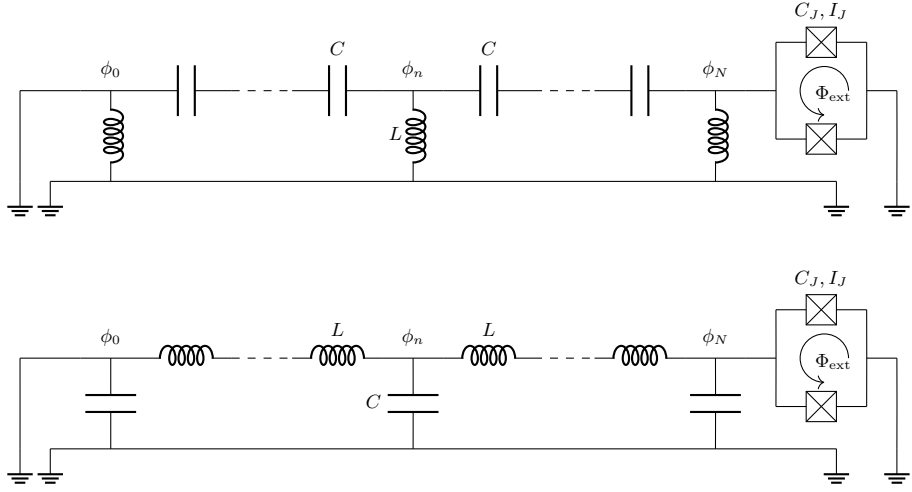
\begin{figure*}[t]
\centering
\begin{circuitikz}[scale=0.8, transform shape]
    \draw (1.5,-1.5) to[short] (14.5,-1.5);
    \draw (1.5,-1.5) node[ground]{};
    \draw (14.5,-1.5) node[ground]{};
    \draw (1,0) to[short] (1,-1.5) node[ground]{};
    \draw (1,0) to[short] (2,0);
    \draw (2,0) to[short, l=$\phi_0$] (3,0);
    \draw (2.5,0) to[L] (2.5,-1.5);
    \draw (3,0) to[C] (4.5,0);
    \draw[dashed] (4.5,0) -- (5.5,0);
    \draw (5.5,0) to[C, l=$C$] (7,0);
    \draw (7,0) to[short, l=$\phi_n$] (8,0);
    \draw (7.5,0) to[L, l_=$L$] (7.5,-1.5);
    \draw (8,0) to[C, l=$C$] (9.5,0);
    \draw[dashed] (9.5,0) -- (10.5,0);
    \draw (10.5,0) to[C] (12,0);
    \draw (12,0) to[short, l=$\phi_N$] (13,0);
    \draw (12.5,0) to[L] (12.5,-1.5);
    \draw (13,0) to[short] (13.5,0);
    \draw (13.5,-0.8) to[short] (13.5,0.8);
    \draw (13.5,-0.8) to[short] (14,-0.8);
    \draw (13.5,0.8) to[short] (14,0.8);
    \draw (14,0.55) rectangle (14.5,1.05);
    \draw (14,0.55) -- (14.5,1.05);
    \draw (14,1.05) -- (14.5,0.55);
    \node[above] at (14.25,1.05) {$C_J, I_J$};
    \draw (14,-0.55) rectangle (14.5,-1.05);
    \draw (14,-0.55) -- (14.5,-1.05);
    \draw (14,-1.05) -- (14.5,-0.55);
    \draw (14.5,-0.8) to[short] (15,-0.8);
    \draw (14.5,0.8) to[short] (15,0.8);
    \draw (15,-0.8) to[short] (15,0.8);
    \draw (15,0) to[short] (15.5,0);
    \draw (15.5,0) to[short] (15.5,-1.5) node[ground]{};
    \node at (14.45, -0.05) {$\Phi_{\text{ext}}$};
    \draw[->, thin] (14.7, 0) arc (0:270:0.4);
\end{circuitikz}

\vspace{0.6cm}

\begin{circuitikz}[scale=0.8, transform shape]
    \draw (1.5,-1.5) to[short] (14.5,-1.5);
    \draw (1.5,-1.5) node[ground]{};
    \draw (14.5,-1.5) node[ground]{};
    \draw (1,0) to[short] (1,-1.5) node[ground]{};
    \draw (1,0) to[short] (2,0);
    \draw (2,0) to[short, l=$\phi_0$] (3,0);
    \draw (2.5,0) to[C] (2.5,-1.5);
    \draw (3,0) to[L] (4.5,0);
    \draw[dashed] (4.5,0) -- (5.5,0);
    \draw (5.5,0) to[L, l=$L$] (7,0);
    \draw (7,0) to[short, l=$\phi_n$] (8,0);
    \draw (7.5,0) to[C, l_=$C$] (7.5,-1.5);
    \draw (8,0) to[L, l=$L$] (9.5,0);
    \draw[dashed] (9.5,0) -- (10.5,0);
    \draw (10.5,0) to[L] (12,0);
    \draw (12,0) to[short, l=$\phi_N$] (13,0);
    \draw (12.5,0) to[C] (12.5,-1.5);
    \draw (13,0) to[short] (13.5,0);
    \draw (13.5,-0.8) to[short] (13.5,0.8);
    \draw (13.5,-0.8) to[short] (14,-0.8);
    \draw (13.5,0.8) to[short] (14,0.8);
    \draw (14,0.55) rectangle (14.5,1.05);
    \draw (14,0.55) -- (14.5,1.05);
    \draw (14,1.05) -- (14.5,0.55);
    \node[above] at (14.25,1.05) {$C_J, I_J$};
    \draw (14,-0.55) rectangle (14.5,-1.05);
    \draw (14,-0.55) -- (14.5,-1.05);
    \draw (14,-1.05) -- (14.5,-0.55);
    \draw (14.5,-0.8) to[short] (15,-0.8);
    \draw (14.5,0.8) to[short] (15,0.8);
    \draw (15,-0.8) to[short] (15,0.8);
    \draw (15,0) to[short] (15.5,0);
    \draw (15.5,0) to[short] (15.5,-1.5) node[ground]{};
    \node at (14.45, -0.05) {$\Phi_{\text{ext}}$};
    \draw[->, thin] (14.7, 0) arc (0:270:0.4);
\end{circuitikz}
\caption{Schematics of the transmission lines terminated by a SQUID. Top:
LHST with series capacitors $C$ and shunt inductors $L$ to
ground. Bottom: RHST with series inductors $L$ and shunt
capacitors $C$ -- same SQUID termination in both ($C_J$, $I_J$, external
flux $\Phi_{\rm ext}$).}
\label{fig:circuits}
\end{figure*}

We start by considering the top circuit of Fig.~\ref{fig:circuits} that is a left-handed array: capacitors $C$ connect neighboring nodes $\phi_n$ in series, while inductors $L$ shunt each node to ground. This reverses the roles of $L$ and $C$ with respect to a standard line (bottom circuit of Fig.~\ref{fig:circuits}, the right-handed array), where inductors sit in series and capacitors shunt each node to ground~\cite{Eleftheriades2002, Kozyrev2008, FerreriWilhemCreation}. In both cases the array is terminated, at the last node $\phi_N$, by a SQUID, consisting of two Josephson junctions forming a loop threaded by an external flux $\Phi_{\rm ext}$~\cite{apC2, Wallquist2006, JohanssonCasimir2010}. The SQUID acts as a tunable, nonlinear boundary element with two Josephson junctions, each with capacitance $C_J$ and critical current $I_J$. Modulating $\Phi_{\rm ext}$ in time is what parametrically drives the system and, as we derive below, sources the dynamical Casimir effect at the boundary. In what follows, we derive  the Hamiltonian of the LHST line, together with the transcendental equation that determines the allowed modes of the line.
 The full derivation of the equations of motion from the circuit topology is detailed in Appendix \ref{app:eqs_motion}.

Taking the node fluxes $\phi_n$ as generalized coordinates~\cite{YurkeDenker1984, VoolDevoret2017}, with $\phi_0=0$ at the grounded left end, the Lagrangian of the circuit has two contributions: the bulk of the array and the boundary at the last node, where the SQUID sits. In the bulk we write $\mathcal{L}_L = T - V$. The series capacitors $C$ give energies quadratic in $\dot \phi_n$ and play the role of kinetic energy $T$, while the shunt inductors $L$ give energies quadratic in $\phi_n$ and make up the potential energy $V$. We add the boundary contribution of the SQUID with its junction capacitance $C_J$ and Josephson energy $E_J(t)$ at the last node ($n=N$): 
\begin{eqnarray}
\mathcal L_L &=& \sum_{n=1}^N \Big[ \frac{C}{2}(\dot\phi_n-\dot\phi_{n-1})^2 - \frac{\phi_n^2}{2L} \Big]\nonumber \\ 
&+&  \Big[C_J\dot\phi_N^2 
+ E_J(t)\cos\Big(\frac{2\pi\phi_N}{\Phi_0}\Big) 
\Big] , \label{eq:lagrangian_tv}
\end{eqnarray}
where $\Phi_0=h/2e$ is
the flux quantum. The effective Josephson energy of the SQUID is set by the external flux
$\Phi_{\rm ext}(t)$ threading its loop, so that $E_J(t)\equiv E_J[\Phi_{\rm ext}(t)]$. In the linear regime, where flux fluctuations at the boundary
are small compared to the flux quantum ($\phi_N\ll\Phi_0$), we expand the cosine
to second order~\cite{Wallquist2006, JohanssonCasimir2010, Weissl2015, FerreriWilhemCreation}: 
\begin{equation*} 
E_J(t)\cos\Big(\frac{2\pi\phi_N}{\Phi_0}\Big)\simeq
E_J(t)-\frac{E_J(t)}{2}\Big(\frac{2\pi\phi_N}{\Phi_0}\Big)^2 \to -\frac{\phi_N^2}{2 L_J(t)}. 
\label{eq:LJ}
\end{equation*}
The zeroth-order term $E_J(t)$ does not depend
on $\phi_N$ and therefore does not contribute to the equations of motion, so we just
drop it. The remaining quadratic term is that of a linear, time-dependent inductance
$L_J(t)=\Phi_0^2/(4\pi^2E_J(t))$. 

We modulate the Josephson
energy harmonically around its static value, $E_J(t)=E_0[1+\eta\cos(\Omega t)]$ where $\Omega$ is the pump frequency and $\eta\ll1$ the relative modulation
amplitude~\cite{JohanssonCasimir2010, Lombardo2016, FerreriWilhemSimulation}. Since $1/L_J(t)\propto E_J(t)$, the inductance
inherits the same modulation around the static Josephson inductance
$L_J\equiv\Phi_0^2/(4\pi^2E_0)$, and the Lagrangian of the LHST line reads

\begin{align}
\mathcal L_L &= \sum_{n=1}^N\left[\frac{C}{2}(\dot\phi_n-\dot\phi_{n-1})^2
-\frac{\phi_n^2}{2L}\right] + C_J\dot\phi_N^2 \nonumber\\
&\quad -\frac{\phi_N^2}{2L_J}
-\frac{\eta\cos(\Omega t)}{2L_J}\phi_N^2. \label{eq:lagrangian_explicit}
\end{align}

Alternatively, Eq.~\eqref{eq:lagrangian_explicit} can be
written in matrix form,
\begin{equation}
\mathcal L_L = \frac{C}{2}\dot{\bar\phi}^T\mathbb A\,\dot{\bar\phi}
-\frac{1}{2L}\bar\phi^T\mathbb B\,\bar\phi
-\frac{\eta\cos(\Omega t)}{2L_J}\phi_N^2, \label{eq:4}
\end{equation}
where $\bar\phi=(\phi_1,\dots,\phi_N)^T$ and $\mathbb A$, $\mathbb B$ are the
dimensionless capacitance and inductance matrices given in
Eqs.~\eqref{eq:A-LH}-\eqref{eq:B-LH} of Appendix~\ref{app:eqs_motion}. The SQUID enters only through their last
diagonal elements, $1+2C_J/C$ in $\mathbb A$ and $1+L/L_J$ in $\mathbb B$. Since
the capacitors are in series and the inductors shunt each node to ground,
$\mathbb A$ is tridiagonal and $\mathbb B$ diagonal, the reverse of the RHST line.

The normal modes of the static part are obtained by solving the generalized eigenvalue problem $\mathbb B\bar u_i=\lambda_i\mathbb A\bar u_i$. Collecting the eigenvectors \(\bar u_i\) as the columns of the modal matrix \(\mathbb U\), and choosing them to be orthonormal with respect to the metric defined by \(\mathbb A\), the generalized eigenvalue problem implies the relations \(\mathbb U^T\mathbb A\mathbb U=\mathbb I,\text{ and }  \mathbb U^T\mathbb B\mathbb U=\mathbf\Lambda, \) where \(\mathbf\Lambda=\operatorname{diag}(\lambda_1,\ldots,\lambda_N)\). These relations allow the Lagrangian to be diagonalized through the modal transformation, reducing its static part to that of \(N\) uncoupled harmonic oscillators.

Motivated by the grounded boundary condition at $n=0$, we propose the ansatz $u_i^n=(1/N_i^L)\sin(k_in\Delta x)$ for the spatial eigenvectors, where $\Delta x$ is the length of a unit cell, so that $n\Delta x$
is the position of node $n$. This yields the LHST dispersion relation:

\begin{equation}
\lambda_i=\frac{1}{4\sin^2\!\left(\frac{k_i\Delta x}{2}\right)}. \label{eq:8}
\end{equation}
Unlike a standard right-handed line, the eigenvalues scale as the inverse square of the sine of the wavevector; we analyze the physical consequences of this anomalous dispersion and its spectral bounds in detail below.

The allowed wavevectors $k_i$ are fixed by the boundary condition at the SQUID end ($n = N$ from the generalized eigenvalue equations), which yields the transcendental equation:
\begin{equation}
\left(1+\frac{L}{L_J}\right)4\sin^2\!\left(\frac{k_i\Delta x}{2}\right)
= 1+\frac{2C_J}{C}-\frac{\sin(k_i(N-1)\Delta x)}{\sin(k_iN\Delta x)}. \label{eq:9}
\end{equation}

Since $\mathbb B$ is diagonal, $\mathbb U^T\mathbb B\,\mathbb U=\Lambda$ gives the normalization constant of each mode:

\begin{equation*}
N_i^L=\sqrt{\frac{\sum_{n=1}^{N-1}\sin^2(k_in\Delta x)
+\left(1+\frac{L}{L_J}\right)\sin^2(k_iN\Delta x)}{\lambda_i}}. \label{eq:10}
\end{equation*}

Besides fixing the allowed $k_i$ through the transcendental equation, the SQUID enters the
normalization only through the weight $1+L/L_J$ of the last node.

We expand the node fluxes in normal modes $\phi_n = \sum_i u_i^n q_i$, thereby mapping the node fluxes $\phi_n$ onto the generalized coordinates $q_i$. Defining the normal mode frequencies as $\omega_i^2 \equiv \lambda_i/LC$, the
Lagrangian becomes: 
\begin{equation*}
    \mathcal{L}_L = \sum_i \left(\frac{C}{2}\dot{q}_i^2 - \frac{C\omega_i^2}{2}q_i^2\right) - \frac{\eta\cos(\Omega t)}{2L_J}\left(\sum_j u_j^N q_j\right)^2.
\end{equation*}

The static part describes a set of harmonic oscillators with effective mass $C$ and frequencies $\omega_i$, coupled only through the modulated boundary term. 
The momentum conjugate to $q_i$ is $p_i = C\dot{q}_i$ and the Legendre transformation gives $H_L = H_0 + H_I(t)$ with:
\begin{align*}
 &  H_L = \sum_i \left(\frac{p_i^2}{2C} + \frac{C\omega_i^2}{2}q_i^2 \right) + \frac{\eta\cos(\Omega t)}{2L_J}\left(\sum_j u_j^N q_j\right)^2 
\end{align*}
where $H_0$ is the sum of free oscillators and  $H_I(t)$ denotes the time-dependent term of the Hamiltonian.

\begin{table*}[t]
\centering
\renewcommand{\arraystretch}{2.0}
\footnotesize
\begin{tabular}{lcc}
\hline\hline
 & LHST & RHST \\
\hline
Capacitance matrix $\mathbb A$ & tridiagonal (series $C$) & diagonal (shunt $C$) \\
Inductance matrix $\mathbb B$ & diagonal (shunt $L$) & tridiagonal (series $L$) \\
Dispersion $\lambda_i$ (with $\omega_i^2,\nu_i^2\equiv\lambda_i/LC$) &
$\dfrac{1}{4\sin^2(k_i\Delta x/2)}$ &
$4\sin^2(p_i\Delta x/2)$ \\
Normalization $N_i$ &
$\sqrt{\dfrac{\sum_{n=1}^{N-1}\sin^2(k_in\Delta x)+\left(1+\frac{L}{L_J}\right)\sin^2(k_iN\Delta x)}{\lambda_i}}$ &
$\sqrt{\sum_{n=1}^{N-1}\sin^2(p_in\Delta x)+\left(1+\frac{2C_J}{C}\right)\sin^2(p_iN\Delta x)}$ \\
Coupling $g_{ij}$ &
$\alpha\sqrt{LC}\,(\lambda_i\lambda_j)^{1/4}\,\dfrac{\sin(k_iN\Delta x)\sin(k_jN\Delta x)}{\sqrt{S_i^LS_j^L}}$ &
$\dfrac{\alpha\sqrt{LC}}{(\lambda_i\lambda_j)^{1/4}}\,\dfrac{\sin(p_iN\Delta x)\sin(p_jN\Delta x)}{\sqrt{S_i^RS_j^R}}$ \\
\hline\hline
\end{tabular}
\caption{Summary of the key expressions derived for the LHST and
RHST lines. The coupling row is written in terms of
sums that do not depend explicitly on $\lambda_i$
$S_i^L\equiv(N_i^L)^2\lambda_i$ and $S_i^R\equiv(N_i^R)^2$. Both
couplings share the same overall $\sqrt{LC}$ scale -- required on
dimensional grounds -- and differ only in the sign of the power of the
dimensionless eigenvalue $\lambda_i$: $g^L\propto(\lambda_i\lambda_j)^{1/4}$
versus $g^R\propto(\lambda_i\lambda_j)^{-1/4}$, i.e. opposite dependence
on mode frequency at fixed circuit parameters.}
\label{tab:summary}
\end{table*}

Quantization follows by promoting $q_i,p_i$ to operators~\cite{VoolDevoret2017, Blais2021}, which yields the 
 Hamiltonian of the driven system:
\begin{align}
\hat H_L/\hbar &= \sum_i\omega_i\left(\hat a_i^\dagger\hat a_i+\tfrac12\right)
+\sum_{ij}g_{ij}^L\cos(\Omega t)\big(\hat a_i\hat a_j+\hat a_i\hat a_j^\dagger \nonumber\\
&\qquad +\hat a_i^\dagger\hat a_j+\hat a_i^\dagger\hat a_j^\dagger\big), \label{eq:Hleft}
\end{align}
with the coupling constants
\begin{equation}
g_{ij}^L=\alpha\,\frac{u_i^N u_j^N}{\sqrt{\omega_i\omega_j}}=\alpha\,\frac{\sin(k_iN\Delta x)\sin(k_jN\Delta x)}{N_i^LN_j^L\sqrt{\omega_i\omega_j}}
 \label{eq:16}
\end{equation}
and $ \alpha\equiv\frac{\eta}{4CL_J}$. Note that the coupling factorizes: each mode enters only through its amplitude $u_i^N$ at the SQUID node weighted by $\omega_i^{-1/2}$. In particular, $(g_{ij}^L)^2=g_{ii}^Lg_{jj}^L$ so that the cross-coupling between two modes is fixed by their single-mode couplings.

\subsection{Right-handed dual}

Exchanging series and shunt elements, the RHST line  derivation proceeds in the same way, with the structure of the matrices reversed: $\mathbb A$ is diagonal and $\mathbb B$ is tridiagonal, while the SQUID still enters through the last diagonal elements $1+2C_J/C$ in $\mathbb A$ and $1+L/L_J$ in $\mathbb B$. We denote the frequencies and wavevectors by 
$\nu_i$ and $p_i$. The explicit matrices and motion equations are given in 
Appendix \ref{app:eqs_motion} and the main results are summarized in Table~\ref{tab:summary}. 

Solving the corresponding eigenvalue problem gives the standard RH dispersion relation $\lambda_i = 4\sin^2\left(p_i\Delta x/2\right)$~\cite{EggerWilhem2013, FerreriWilhemSimulation}, with eigenfrequencies $\nu_i^2=\lambda_i/LC$ and spatial eigenvectors $u_i^n = (1/N_i^R) \sin(p_i n\Delta x)$.

The allowed wavevectors $p_i$ are fixed by the boundary condition at the SQUID ($n=N$):
\begin{equation*}
1 + \frac{L}{L_{J}} - \frac{\sin(p_i(N-1)\Delta x)}{\sin(p_iN\Delta x)}
    = 4\sin^2\left(\frac{p_i\Delta x}{2}\right)\left(1+\frac{2C_J}{C}\right).\end{equation*}

Since $\mathbb A$ is diagonal, the normalization follows directly from
$\mathbb U^T\mathbb A\,\mathbb U=\mathbb I$,
     \begin{equation*}
    N_i^R = \sqrt{\sum_{n=1}^{N-1}\sin^2(p_i n\Delta x) + \left(1+\frac{2C_J}{C}\right)\sin^2(p_i N\Delta x)}.
\end{equation*}
and the coupling has the same form as in the LH line,
\begin{equation}
    g_{ij}^R =  \alpha\frac{\sin(p_i N\Delta x)\sin(p_j N\Delta x)}{N_i^R N_j^R \sqrt{\nu_i \nu_j}}.
    \label{giiR}
\end{equation}

The difference is that the normalization factor contains no factor of $\lambda_i$. 

In the continuous limit ($N \to \infty$, $\Delta x\to 0$, $C\to0$ and $L\to0$) with $L_{\text{tot}} = NL$ and $C_{\text{tot}} = NC$ fixed, the equation for the allowed wavevectors reduces to that of a continuous line of length $d = N \Delta x$ terminated by a SQUID~\cite{Lombardo2016, Lombardo2018},
\begin{equation}
-p_i d\,\cot(p_i d)=\frac{L_{\text{tot}}}{L_J}-\frac{2C_J}{C_{\text{tot}}}(p_i d)^2,
\label{eq:continuous}
\end{equation}
 with the linear bulk dispersion $\nu_i = v p_i$, $v = d/\sqrt{L_{\text{tot}}C_{\text{tot}}}$.

\subsection{Transmission line spectra and coupling strengths}
\label{sec:frecsycoupls}

\begin{figure}[htbp]
    \centering \includegraphics[width=0.75\linewidth]{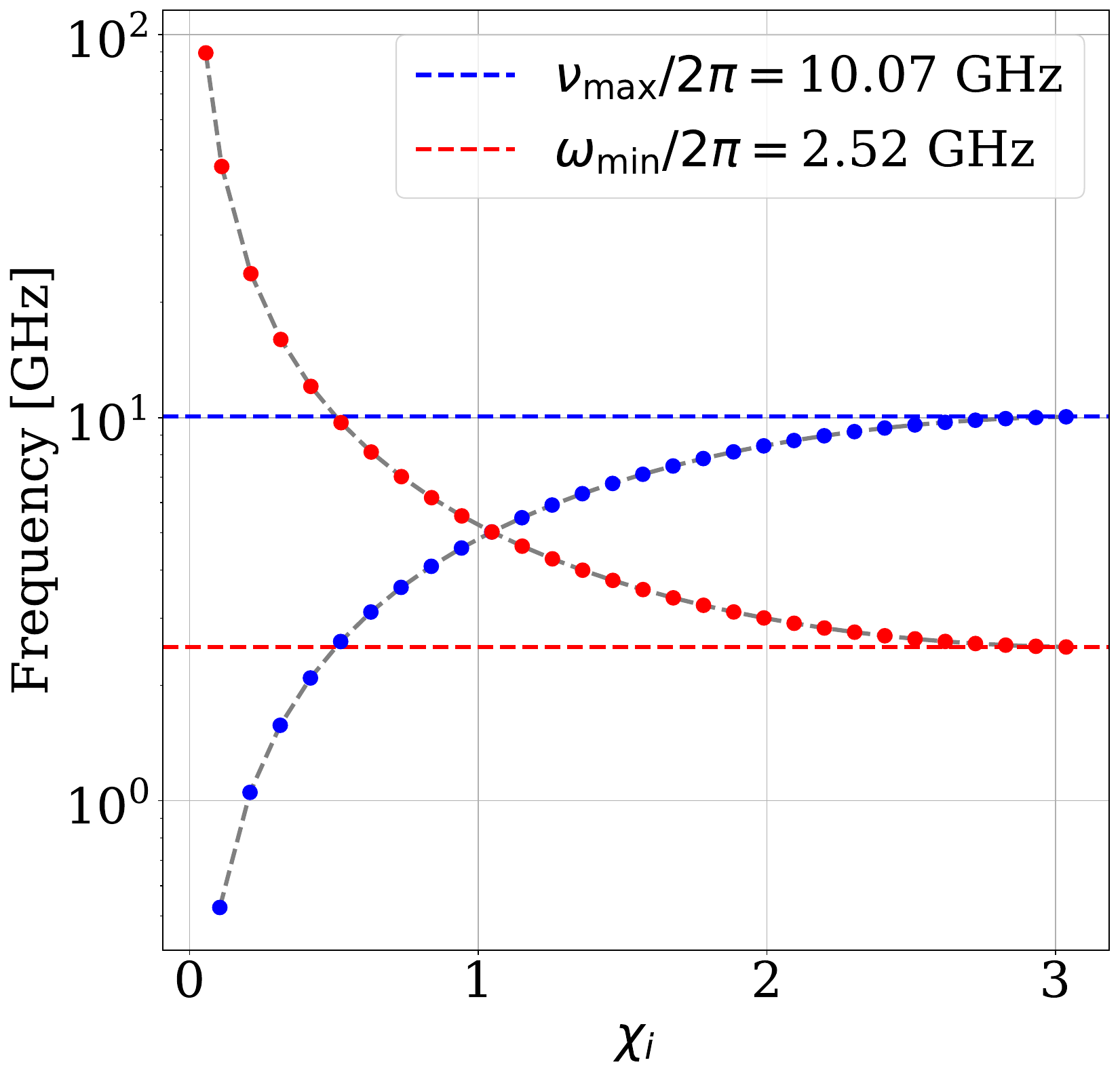}
    \caption{Eigenfrequencies of the LHST ($\omega_i/2\pi$, red circles) and RHST ($\nu_i/2\pi$, blue circles) lines as a function of the dimensionless wavevector $\chi_i$ ($\chi_i = k_i\Delta x$ for LHST and $\chi_i = p_i\Delta x$ for RHST). The horizontal dashed lines indicate the upper band edge of the RHST line, $\nu_{\max}/2\pi = 10.07$GHz, and the lower band edge of the LHST line, $\omega_{\min}/2\pi = 2.52$GHz. Parameters used: $\eta=0.01$, $L=1\text{nH}$, $C=1\text{pF}$, $I_J=10\mu\text{A}$, $C_J=0.1\text{pF}$, $N=30$.}
    \label{fig:frecuencias}
\end{figure}

Figure~\ref{fig:frecuencias} shows the eigenfrequencies of both lines for $N=30$
  as a function of $\chi_i$, with $\chi_i=k_i\Delta x$ for the LHST line and
  $\chi_i=p_i\Delta x$ for the RHST line.
In the RHST line the frequency grows with $\chi_i$, from $\nu_i\to 0$ at
$\chi_i\to 0$ up to the cutoff $\nu_{\max}=2/\sqrt{LC}$ at $\chi_i\to\pi$,
where the modes pile up. At small $\chi_i$ the dispersion is linear and the
modes are almost equidistant, as in the continuous line of
Eq.~\eqref{eq:continuous}. 
The SQUID does not change this, because
$L_{\text{tot}}/L_J=NL/L_J$ is large~\cite{Lombardo2016}. In this regime a
pump at $\Omega=2\nu_k$ also meets, or nearly meets, the resonance conditions
$\Omega\approx\nu_i\pm\nu_j$ of other pairs, and particle creation spreads
over many modes~\cite{DodonovKlimov1996,  Lombardo2016}.

The LHST line behaves the opposite way ~\cite{EggerWilhem2013, Wang2019}. Its frequency decreases with
$\chi_i$, from $\omega_i\to\infty$ at $\chi_i\to 0$ down to
$\omega_{\min}=1/(2\sqrt{LC})$ at $\chi_i\to\pi$, where the modes pile up.
At small $\chi_i$ we have $\omega_i\simeq 1/(\chi_i\sqrt{LC})$, so the
highest modes are far apart and not equidistant. A pump tuned to one of them
in general meets no other resonance condition. A purely LH line has no
continuum limit: in real devices, parasitic series inductances and shunt
capacitances dominate at very high frequencies~\cite{LaiItohCaloz2004,Caloz2004, CalozItohBook}.

The number of cells changes the two spectra in different ways. Both
transcendental equations have $N$ solutions in $\chi_i\in(0,\pi)$, spaced by
about $\pi/N$. In the RHST line, a larger $N$ fills the fixed band
$(0,\nu_{\max}]$ more densely and lowers the fundamental frequency, $\nu_{\min}\propto 1/N$. In the LHST line, a larger $N$ fills the band $\left[\omega_{\min},\infty\right)$ more densely, while the lowest wavevector $\chi_{\min}\sim\pi/N$ sets the highest frequency, which grows as $\omega_{\max}\sim N$.

\begin{figure}[htbp]
    \centering
    \begin{subfigure}{0.45\columnwidth}
        \centering
        \includegraphics[width=\linewidth]{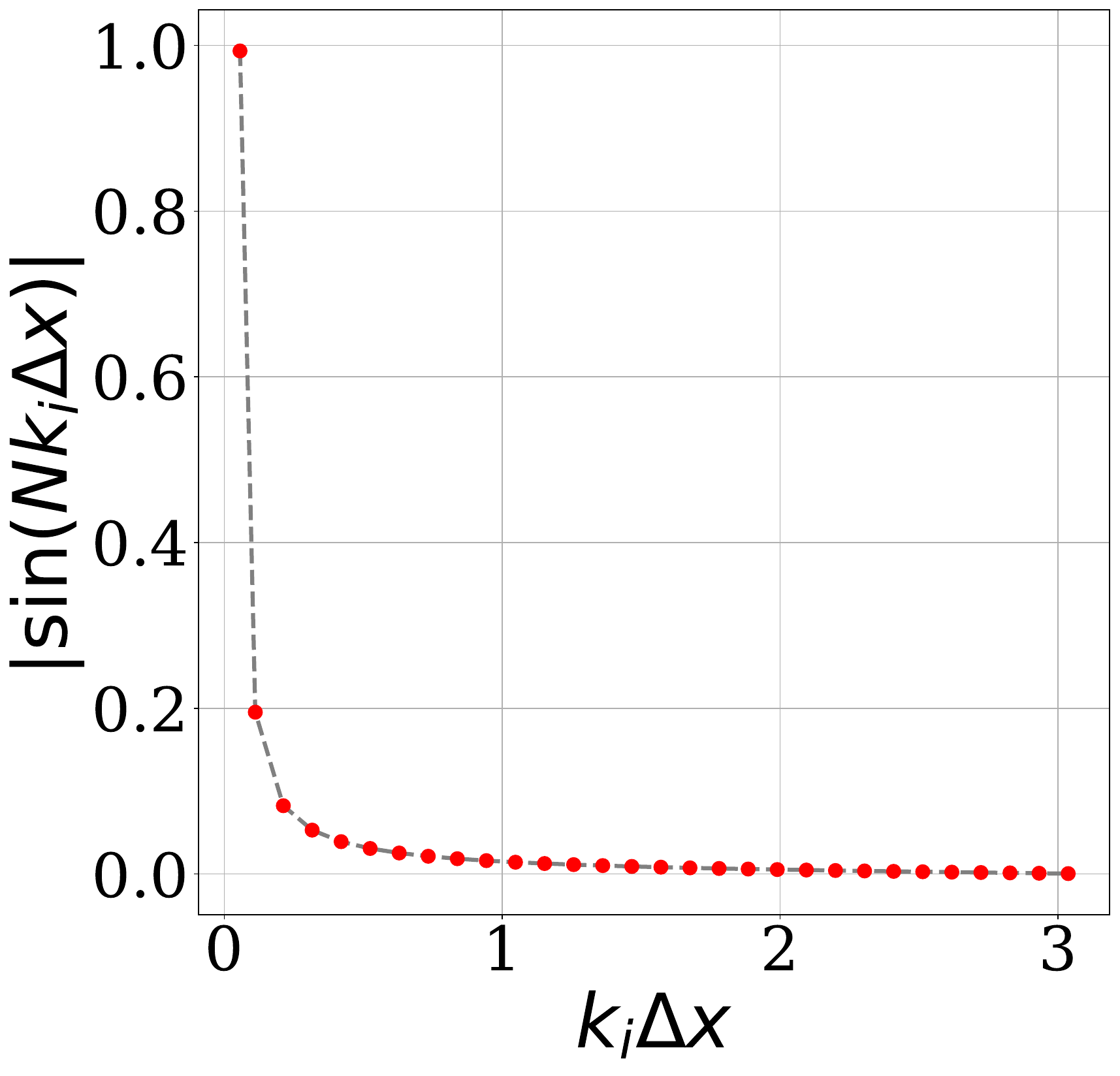}
        \label{fig:subfig3}
    \end{subfigure}
    \hfill
    \begin{subfigure}{0.45\columnwidth}
        \centering
        \includegraphics[width=\linewidth]{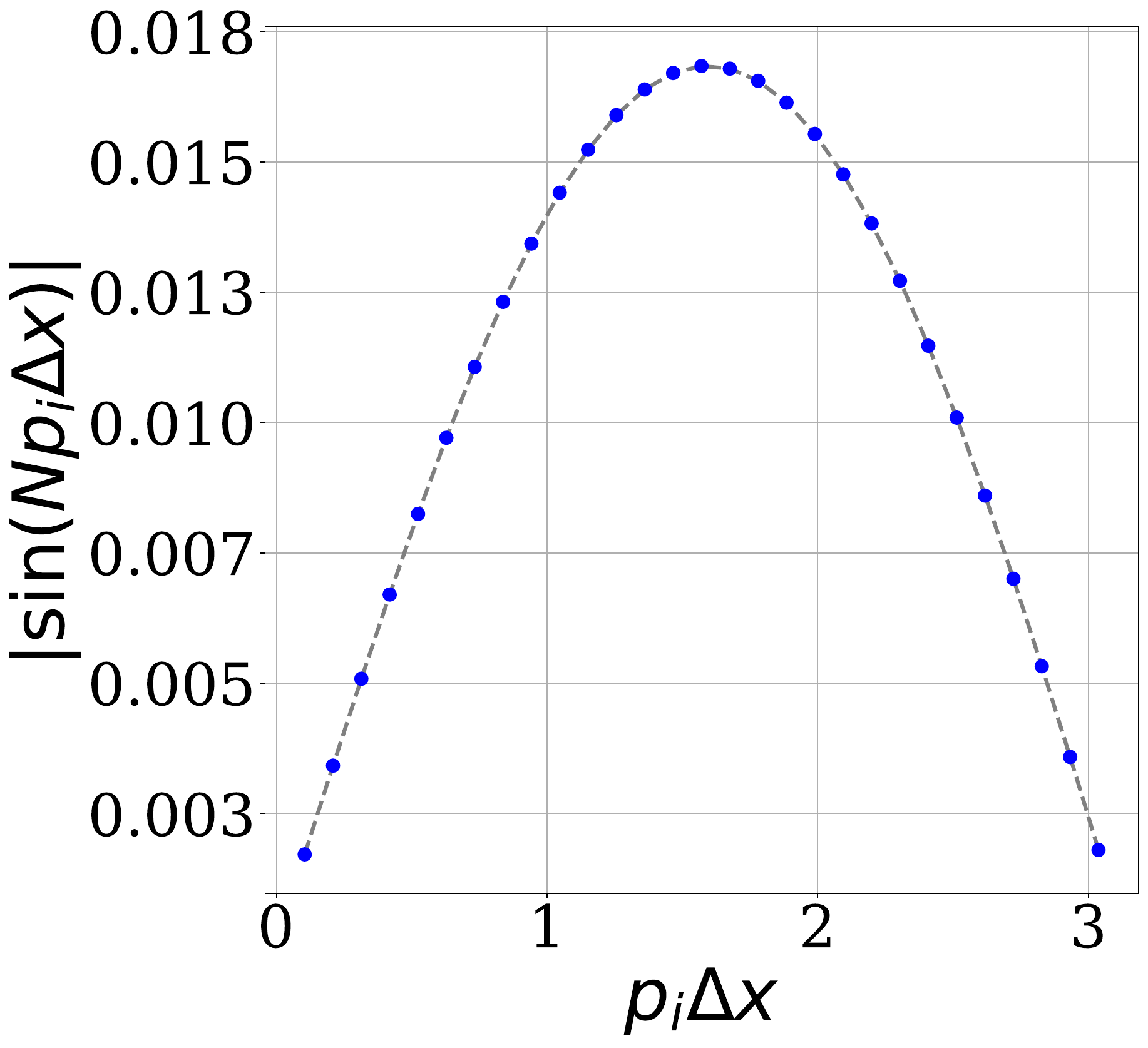}
        \label{fig:subfig4}
    \end{subfigure}
    \caption{Boundary amplitudes as a function of the dimensionless wavevector. The difference between these boundary terms arises from the geometrical differences in the transcendental equations. Circuit parameters match those of Fig.~\ref{fig:frecuencias}.}
    \label{fig:gvsbordes}
\end{figure}

From Table \ref{tab:summary}, it can be observed that both the right and left couplings exhibit the same dependence on the factor $4\sin^2(\chi_i/2)$ through opposite dependence of $\lambda_i$. One might therefore expect their behavior to be identical as the dimensionless wavevector varies. However, this is not the case, as the coupling also depends on the boundary amplitude of the corresponding eigenfunction. Fig. \ref{fig:gvsbordes} illustrates the dependence of the boundary amplitude on the dimensionless wavevector for the LHST and RHST. While the left boundary amplitude spans the entire allowed range $(0,1)$, the right boundary amplitude remains considerably more constrained. As discussed below, this difference allows the left coupling to exhibit a significantly broader range of values, providing additional flexibility that can be exploited to our advantage. These distinct behaviors of the boundary terms are primarily determined by the geometrical differences between the corresponding transcendental equations. The robustness of the left advantage against variations in the boundary parameters $C_J$, $L_J$ is discussed in Appendix \ref{app:couplings}
.

\section{Particle creation}
\label{sec:singleresonance}

Modulating the boundary of a resonator creates particles from vacuum when the
pump is in parametric resonance with the field modes. Particles are then
created in pairs, either both in the same mode or one in each of two modes,
and their number grows exponentially in time as long as no other mode takes
part in the process. Other resonances only move particles from one mode to another. Which processes a given pump drives therefore depends on the spectrum
of the line, and how fast particles are created depends on how strongly each
mode couples to the SQUID (Sec.~\ref{sec:frecsycoupls}).

\subsection{One mode resonance}
\label{1res}

We first consider a single-mode resonance, $\Omega\approx2\omega_k$, for a given mode $k$, and define the detuning $\Delta_k\equiv\Omega-2\omega_k$. The only resonant terms are the diagonal ones, $\hat a_k^{\dagger2}e^{-i\Omega t}$ and its Hermitian conjugate, which appear once in the double sum and therefore carry strength $g_{kk}/2$. This holds for $|\Delta_k|,\,g_{kk}\ll\omega_k$ and provided no other pair satisfies $\omega_i+\omega_j\approx\Omega$ within $g_{ij}$. In the Schrödinger picture, the RWA Hamiltonian reads
\begin{equation}
\hat H_{\rm RWA}/\hbar=\omega_k\hat a_k^\dagger\hat a_k+\frac{g_{kk}}{2}\left(\hat a_k^{\dagger2}e^{-i\Omega t}+\hat a_k^{2}e^{i\Omega t}\right).
\label{eq:HRWAonemode}
\end{equation}
Both terms become time independent in a frame rotating at half the pump frequency, $\hat{\mathcal U}=\exp[-i(\Omega/2)\hat a_k^\dagger\hat a_k t]$: the transformation turns $\hat a_k^{\dagger2}e^{-i\Omega t}$ into $\hat a_k^{\dagger2}$ and shifts the free term by $-\hbar(\Omega/2)\hat a_k^\dagger\hat a_k$, leaving the residual frequency $\omega_k-\Omega/2=-\Delta_k/2$. The effective Hamiltonian is

\begin{equation}
    \hat{H}_{\text{eff}}/\hbar=-\frac{\Delta_k}{2}\hat{a}_k^{\dagger}\hat{a}_k+\frac{g_{kk}}{2}(\hat{a}_k^{2}+\hat{a}_k^{\dagger2}).
    \label{eq:Honemode}
\end{equation}
Since $\hat{H}_{\rm{eff}}$ is time independent, the Heisenberg equation $\dot{\hat a}_k = (i/\hbar)[\hat H_{\rm eff},\hat a_k]$ in the rotating frame gives
\begin{equation} \dot{\hat{a}}_k=i\frac{\Delta_k}{2}\hat{a}_k-i g_{kk}\hat{a}_k^{\dagger},
\end{equation}
together with its Hermitian conjugate, gives $\ddot{\hat{a}}_k=\Gamma_k^2\hat{a}_k$ with
$\Gamma_k^2=g_{kk}^2-\left(\frac{\Delta_k}{2}\right)^2$, whose solution is
\begin{eqnarray}
    \hat{a}_k(t) &=& \hat{a}_k(0)\left[\cosh(\Gamma_k t)+i\frac{\Delta_k}{2\Gamma_k}\sinh(\Gamma_k t)\right] \nonumber\\
    && -\,i\frac{g_{kk}}{\Gamma_k}\hat{a}_k^{\dagger}(0)\sinh(\Gamma_k t).
\end{eqnarray}
For the vacuum initial state the mean number of particles created is 
\begin{equation}  \langle\hat{N}_k\rangle=\frac{g_{kk}^2}{\Gamma_k^2}\sinh^2(\Gamma_k t),
\end{equation}
identical in the laboratory frame, since $\hat{\mathcal{U}}$ commutes with $\hat{a}_k^{\dagger}\hat{a}_k$ (Appendix~\ref{app:rwa_and_rot}). Particle creation grows exponentially only if $\Gamma_k$ is real, i.e.,
\begin{equation}
|g_{kk}|>\left|\frac{\Delta_k}{2}\right|;
\end{equation}
below this threshold $\Gamma_k$ is imaginary and $\langle\hat{N}_k\rangle$ oscillates and remains bounded~\cite{Lombardo2018, LouysSanso2025}. On resonance, $\Delta_k=0$, the mode evolves into a squeezed vacuum with squeezing parameter $r=g_{kk}t$, so that $\langle\hat{N}_k\rangle=\sinh^2(g_{kk}t)$~\cite{ScullyZubairy1997, FerreriWilhemCreation}.

\begin{figure*}[htbp]
    \centering
    
    \begin{subfigure}{0.95\linewidth} 
        \centering
        \includegraphics[width=\linewidth]{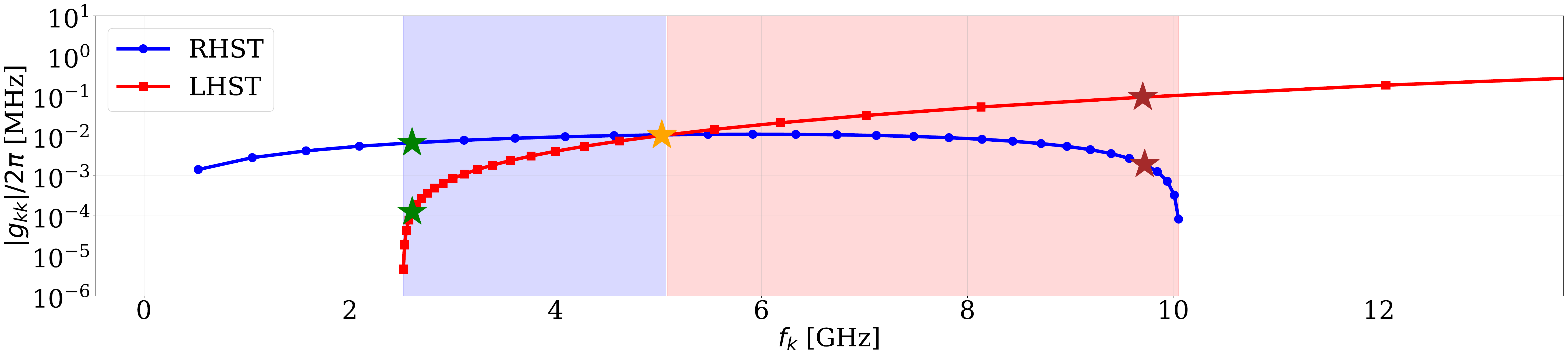}
        \label{fig:arriba}
    \end{subfigure}
    
    \vspace{1.em} 
    
    \begin{subfigure}{0.3\linewidth}
        \centering
        \includegraphics[width=\linewidth]{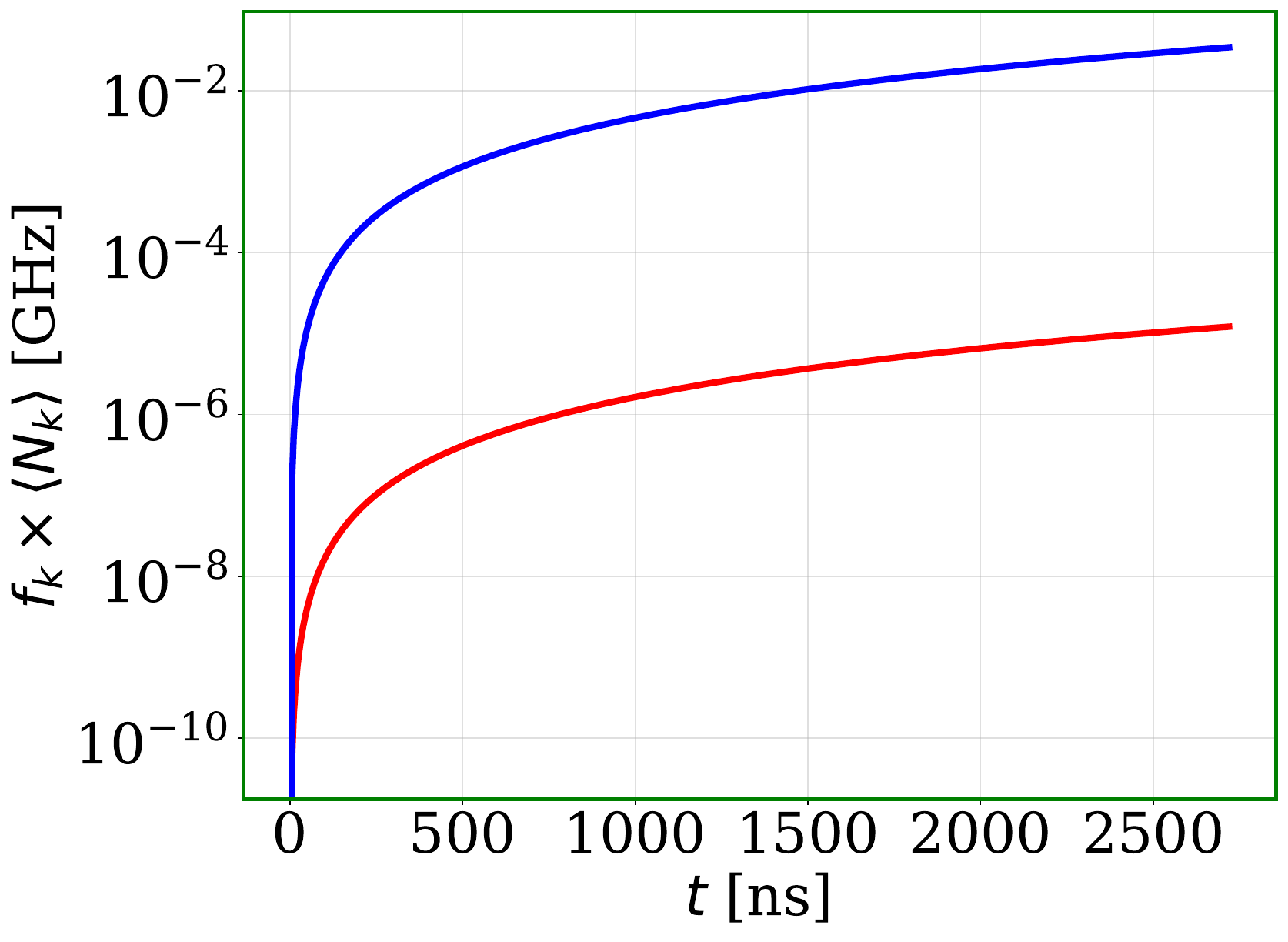}
        \label{fig:evol_right}
    \end{subfigure}\hfill
    \begin{subfigure}{0.3\linewidth}
        \centering
        \includegraphics[width=\linewidth]{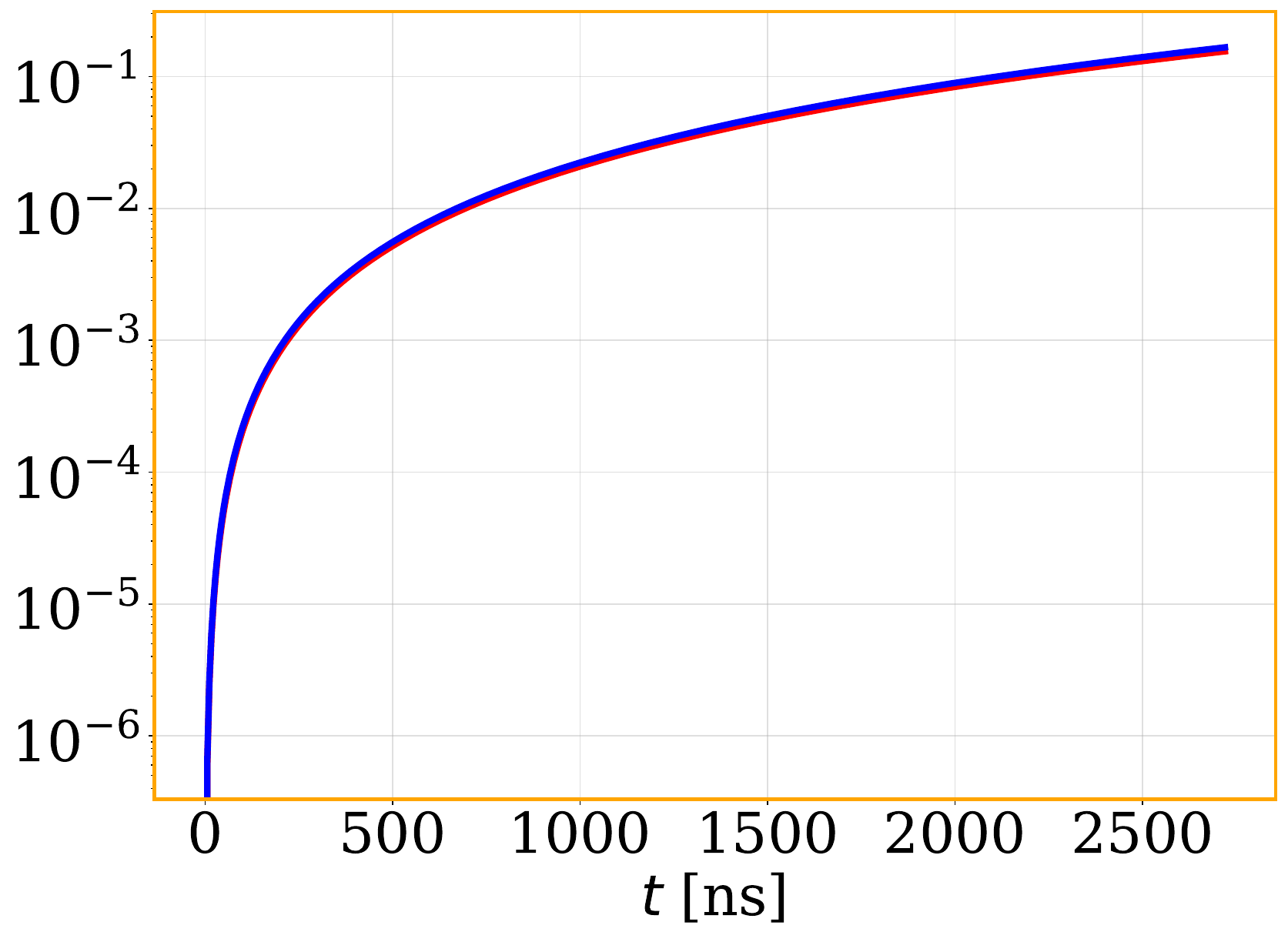}
        \label{fig:evol_igual}
    \end{subfigure}\hfill
    \begin{subfigure}{0.3\linewidth}
        \centering
        \includegraphics[width=\linewidth]{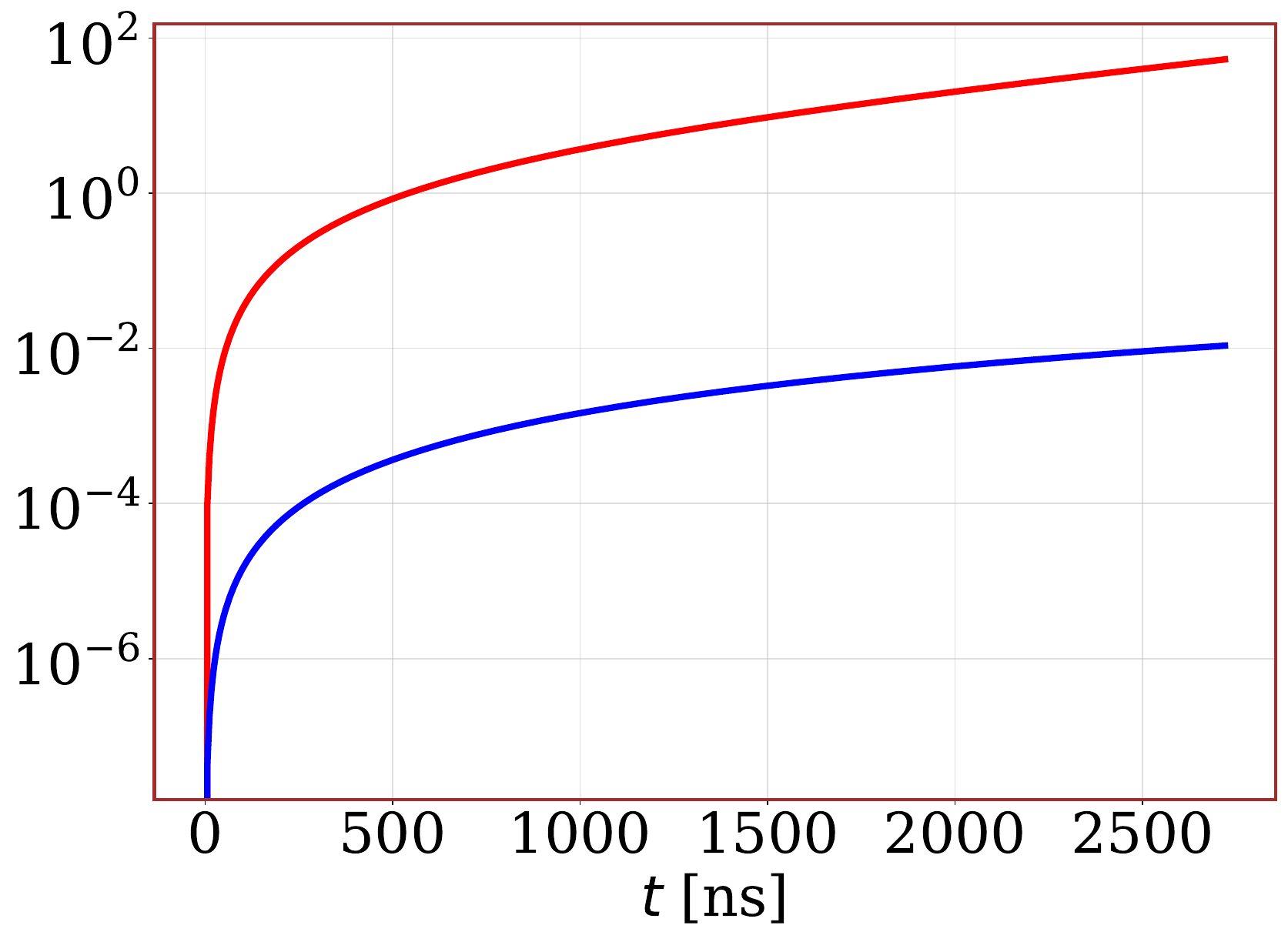}
        \label{fig:evol_left}
    \end{subfigure}
    
    \caption{Top panel: Single-mode couplings $|g_{kk}|/2\pi$ of both lines as a function of the mode frequency $f_k$. The frequency is $f_k = \nu_k/2\pi$ for the RHST and $f_k = \omega_k/2\pi$ for the LHST. 
    The red (blue) region marks where $g^L_{kk} > g^R_{kk}$
($g^L_{kk} < g^R_{kk}$). The colored star markers correspond to the specific operational points analyzed below. 
    Bottom panels: Time evolution of the energy of the target mode, $f_k\langle
N_k\rangle$. The border colors of these plots match the corresponding star markers in the top panel. 
    From left to right, the panels display three distinct operational regimes: the LHST creating fewer particles (left with $\omega_k/2\pi=2.6\text{GHz}$, $g_{kk}^L/2\pi=2\times10^{-4}\text{MHz}$ and $\nu_k/2\pi=2.6\text{GHz}$, $g_{kk}^R/2\pi=7\times10^{-3}\text{MHz}$), both lines creating a similar number (center with $\omega_k/2\pi=\nu_k/2\pi=5\text{GHz}$, $g_{kk}^L/2\pi=g_{kk}^R/2\pi=1\times10^{-2}\text{MHz}$), and the LHST creating many more particles (right with $\omega_k/2\pi=9.7\text{GHz}$, $g_{kk}^L/2\pi=1\times 10^{-1}\text{MHz}$ and $\nu_k/2\pi=9.7\text{GHz}$, $g_{kk}^R/2\pi=2\times 10^{-3}\text{MHz}$).      Circuit parameters match those of Fig.~\ref{fig:frecuencias}.}   \label{fig:amplificacionyatenuacion}
\end{figure*}

The top panel of Fig.~\ref{fig:amplificacionyatenuacion} compares the single-mode couplings of the two lines. For the RH line, $g_{kk}^R$ remains nearly constant across the band (except near its upper band edge), whereas $g_{kk}^L$ increases with frequency.

The LH spectrum also extends to frequencies beyond the upper band edge of the RH line. Near its lower band edge, however, the LH modes have a strongly suppressed amplitude at the SQUID node and consequently couple only weakly to the modulation. The LH line therefore produces substantially fewer particles than the RH line near its lower band edge, a comparable number at intermediate frequencies, and significantly more particles at high frequencies. This behavior is reflected in the bottom panels, which show the energy of the target mode, $E_k/h=\hbar\omega_k\langle\hat{N}_k\rangle/h=(\omega_k/2\pi)\langle\hat{N}_k\rangle=f_k\langle\hat{N}_k\rangle$, at three representative frequencies. Thus, single-mode pumping does not provide a uniform enhancement of the LH line; rather, it redistributes the coupling toward higher frequencies, at the expense of the modes near its lower band edge. As we show next, this suppression can be overcome by pumping two modes at once.

\subsection{Two mode resonance}
\label{2res}

In the single-mode case only the diagonal couplings of $g_{kk}$ survived the RWA. However, in Eq.~\eqref{eq:Hleft} every pair of modes is coupled through the SQUID node, and the off-diagonal terms become resonant when 
the pump frequency satisfies $\Omega \approx \omega_k + \omega_l$ (with $k \neq l$)~\cite{Crocce2001, Lombardo2018}. Particles are then created in pairs, one in mode $k$ and one in mode $l$, so that both modes are equally populated at all times~\cite{Schneider2020, FerreriWilhemSimulation}.

The resonant terms are now $\hat a_k^\dagger\hat a_l^\dagger e^{-i\Omega t}$ and its Hermitian conjugate. Since $k\neq l$, they appear twice in the double sum of Eq.~\eqref{eq:Hleft}, as $(k,l)$ and $(l,k)$, so the pair is driven with strength $g_{kl}$ rather than $g_{kk}/2$. Each mode now needs its own rotating frequency, and the only requirement to make the pair-creation term time independent is $\omega_{{\rm rot},k}+\omega_{{\rm rot},l}=\Omega$. We split the detuning $\Delta_{kl}\equiv\Omega-(\omega_k+\omega_l)$ symmetrically between the two modes, $\omega_{{\rm rot},k}=\omega_k+\Delta_{kl}/2$ and $\omega_{{\rm rot},l}=\omega_l+\Delta_{kl}/2$, i.e., $\hat{\mathcal U} = \exp[-i(\omega_{{\rm rot},k}\hat a_k^\dagger \hat a_k
+ \omega_{{\rm rot},l}\hat a_l^\dagger \hat a_l)t]$, so that each mode is left with a residual frequency $-\Delta_{kl}/2$. The full sequence of steps for this case is written out in Appendix~\ref{app:rwa_and_rot}; the result is

\begin{equation}
\hat{H}_{\text{eff}}/\hbar = -\frac{\Delta_{kl}}{2}\hat{a}_k^\dagger \hat{a}_k - \frac{\Delta_{kl}}{2}\hat{a}_l^\dagger \hat{a}_l + g_{kl}(\hat{a}_k^\dagger \hat{a}_l^\dagger + \hat{a}_k \hat{a}_l).
\label{eq:Htwomode}
\end{equation}
This holds for $|\Delta_{kl}|, g_{kl}\ll\omega_k,\omega_l$ and
assumes no other pair $(i,j)$, including $i=j$, satisfies
$\omega_i+\omega_j\approx\Omega$ or $|\omega_i-\omega_j|\approx\Omega$ within $g_{ij}$.

Proceeding as in the single-mode case, the Heisenberg equations now couple $\hat a_k$ to $\hat a^\dagger_l$,
\begin{equation}
\begin{aligned}
\dot{\hat a}_k &= i\frac{\Delta_{kl}}{2}\hat a_k-ig_{kl}\hat a_l^\dagger,\\
\dot{\hat a}_l^\dagger &= -i\frac{\Delta_{kl}}{2}\hat a_l^\dagger+ig_{kl}\hat a_k,
\end{aligned}
\end{equation}
so that $\ddot{\hat a}_k=\Gamma_{kl}^2\hat a_k$ with $\Gamma_{kl}^2=g_{kl}^2-(\Delta_{kl}/2)^2$. Their solution reads
\begin{eqnarray}
\hat a_k(t)&=&\hat a_k(0)\left[\cosh(\Gamma_{kl}t)+\frac{i\Delta_{kl}}{2\Gamma_{kl}}
\sinh(\Gamma_{kl}t)\right] \nonumber \\
&-&\frac{i g_{kl}}  {\Gamma_{kl}} \hat{a}_l^\dagger(0) \sinh(\Gamma_{kl}t),
    \label{eq:ak(t)}
\end{eqnarray}
and $\hat a_l(t)$ follows from
$k\leftrightarrow l$,
\begin{eqnarray}
    \hat{a}_l(t) &=& \hat{a}_l(0)\left[\cosh(\Gamma_{kl} t) + i\frac{\Delta_{kl}}{2\Gamma_{kl}}\sinh(\Gamma_{kl} t)\right]  \nonumber \\ &-&i\frac{g_{kl}}{\Gamma_{kl}}\hat{a}_k^\dagger(0)\sinh(\Gamma_{kl} t).
      \label{eq:al(t)}
\end{eqnarray}

Starting from vacuum,
\begin{equation}
\langle N_k(t)\rangle=\langle N_l(t)\rangle
=\left(\frac{g_{kl}}{\Gamma_{kl}}\right)^2\sinh^2(\Gamma_{kl}t),
\label{eq:NkNl}
\end{equation}
consistent with $\hat H_{\rm eff}$ conserving $\hat N_k-\hat N_l$. Exponential growth requires $\Gamma_{kl}$ to be real; otherwise $\langle N_k\rangle$
oscillates and remains bounded. Hence, the instability threshold is

\begin{equation}
|g_{kl}| > \left|\frac{\Delta_{kl}}{2}\right|.
\label{resonancecond}
\end{equation}

As anticipated before, the weak particle creation of the LH line near its lower band edge is not a fundamental limitation. Since the coupling factorizes into a product of single-mode factors (Eq.~\eqref{eq:16}), a weakly coupled low-frequency mode $k$ can borrow coupling strength from a strongly coupled high-frequency partner $l$. This is particularly effective in the LH line, whose high-frequency modes have single-mode couplings well above the largest RH ones, whereas in the RH line the single-mode couplings are bounded and pairing gains little. 

\begin{figure}[htbp]
    \centering
    \includegraphics[width=0.85\linewidth]{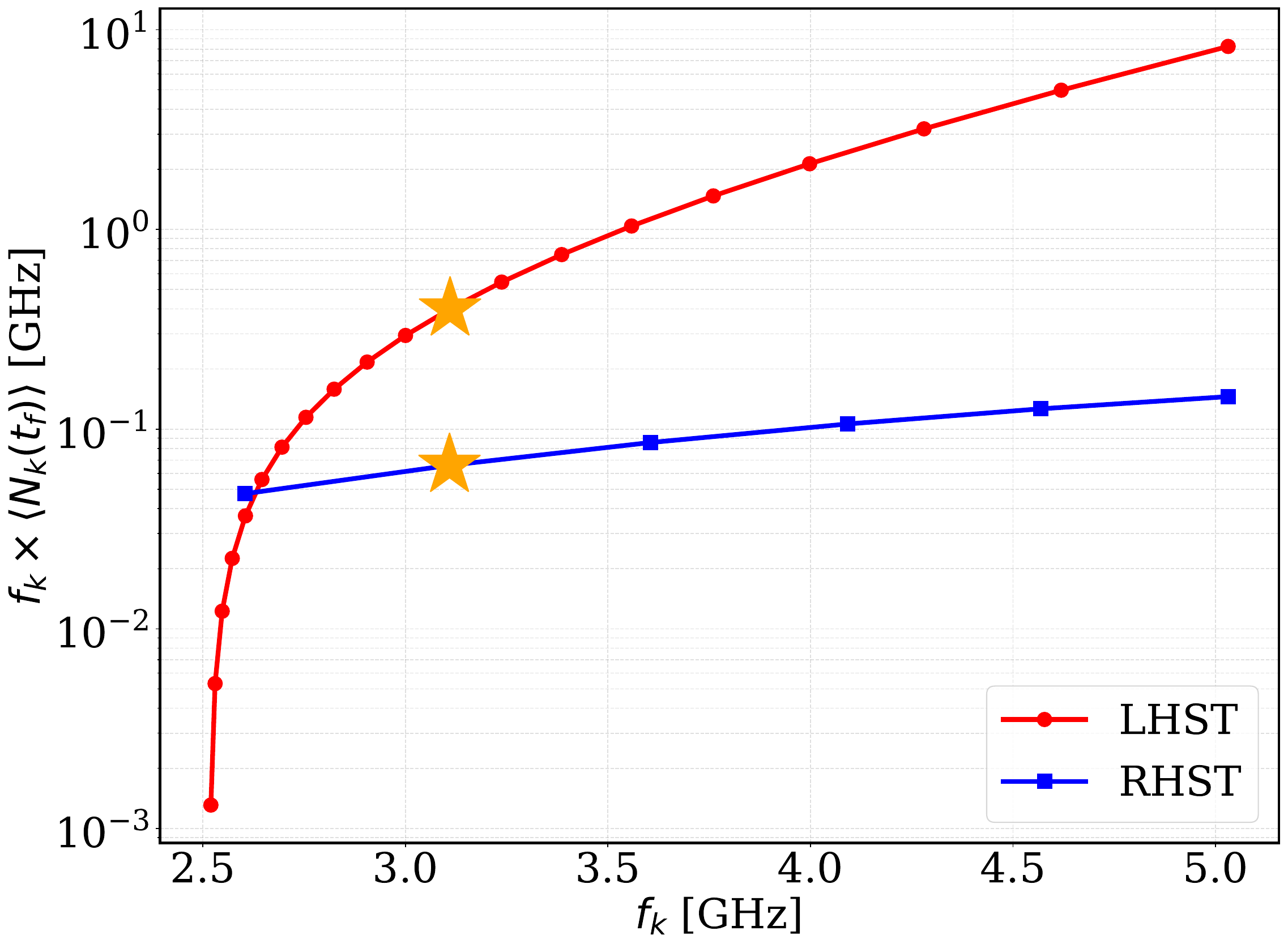}
    \caption{Final energy of the target mode as a function of the target mode frequency $f_k$. The red line corresponds to the final energy, computed as $f_k \times \langle N_k(t_f) \rangle$, for the target mode $k$ in the LHST line utilizing an optimized two mode pumping scheme ($\Omega = \omega_k + \omega_l$). The blue line represents the maximum achievable energy for the identical target frequency $f_k$ in the RHST line, allowing for either one mode or two mode pumping. The star markers highlight a specific example at $f_k = 3.1 \text{GHz}$: the LHST pairs it with an auxiliary mode at $f_l=16.0 \text{GHz}$ ($\Omega = 19.1 \text{GHz}$ and $g_{kl}^L/2\pi=0.022\text{MHz}$), whereas the RHST optimally pairs it with a $f_l=5.9 \text{GHz}$ mode ($\Omega = 9.0 \text{GHz}$ and $g_{kl}^R/2\pi=0.010\text{MHz}$). Parameters are identical to those in Fig. \ref{fig:frecuencias}, evaluated at final time $t_f = 2500 \text{ns}$.}
    \label{fig:ampl_twomode}
\end{figure}

Fig.~\ref{fig:ampl_twomode}  compares both lines at the same target frequency $f_k$, each with its optimal pump. For the LH line (red), the target is paired with the partner $l$ that maximizes the final energy $f_k\langle N_k(t_f)\rangle$ under $\Omega=\omega_k+\omega_l$; for the RH line (blue), we take the best of single-mode pumping and all two-mode pairings. 
The LH target mode gets more energy than the best RH option, about 15 times more at 3.5GHz and about 50 times more at 5GHz, and the gap widens with $f_k$ because $g_{kk}^L$ rises steeply away from the lower band edge, while $g_{kk}^R$ stays nearly flat. Only very close to that edge is $g_{kk}^L$ too small to be compensated for. The cost of this scheme is an equal number of particles in the high-frequency partner.

\subsection{Multiple resonance}
\label{sec:multipleresonance}

In this subsection, we study an effect that comes solely from 
the transcendental equation for the allowed frequencies of the system. Specifically, we analyze how the resonance conditions  $|\omega_i + \omega_j|$ and $|\omega_i - \omega_j|$ evolve as one of the circuit parameters, such as the inductance \(L\), is varied. A first intuition might suggest that the resonance lines associated with both particle-creation and particle-exchange processes~\cite{Crocce2001, Crocce2002, DelGrosso2020} would all scale by the same factor, since every frequency carries the prefactor $1/\sqrt{LC}$. However, the inductance \(L\) also enters the transcendental equation for  \(k_i\), through the SQUID term $L/L_J$. Consequently, the resulting dependence of the resonance conditions on \(L\) is mode-dependent, and the different resonance lines do not shift uniformly~\cite{Lombardo2018, Velasco2022}. This gives rise to crossings between resonance lines.

At these crossings a single pump frequency $\Omega$ satisfies two resonance conditions at once, so it can act on three or four modes. Whether this changes the dynamics depends on whether the two processes share a mode, as we discuss below.
Crossings appear in both lines; the analysis below applies to both arrays but the numerical examples are for the LHST only. These examples use a different SQUID and a shorter line ($I_J = 1\mu$A, $C_J = 0.5$pF, $N = 20$), so their couplings are not directly comparable with those of Fig. ~\ref{fig:amplificacionyatenuacion}.

\begin{figure}[htbp]
    \centering
    \includegraphics[width=0.85\linewidth]{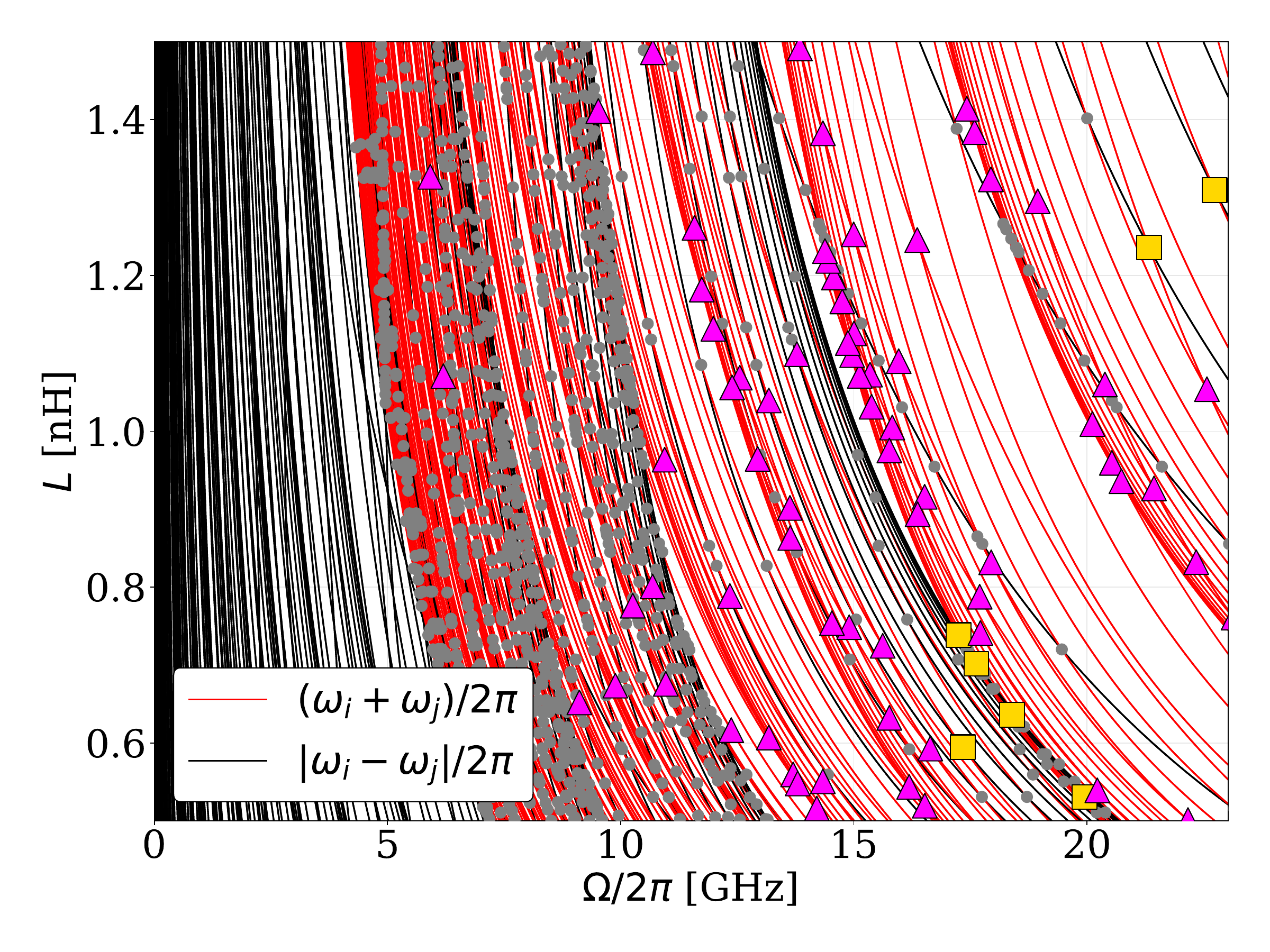}
    \caption{Sweep of the inductance $L$ as a function of the resonance frequencies $\Omega/2\pi$. 
    Red lines denote particle creation resonances $(\omega_i+\omega_j)/2\pi$, while black lines represent particle exchange resonances $|\omega_i-\omega_j|/2\pi$. 
    The markers indicate specific mode crossings where multiple resonance conditions are simultaneously met. 
    The yellow squares mark the useful creation exchange crossings. 
    The purple triangles highlight the creation creation crossings. 
    Gray circles denote creation exchange crossings that do not lead to a three-mode resonance, either because they lack shared indices between the creation and exchange pairs, or because the particle exchange coupling is stronger than the particle creation coupling. Parameters used: $\eta=0.01$, $L=1$nH, $C=1$pF, $I_J=1\mu$A, $C_J=0.5$pF, $N=20$.}
    \label{fig:barridoindL}
\end{figure}

Fig.~\ref{fig:barridoindL} shows these crossings for the LHST with the pump frequency $\Omega/2\pi$ on the horizontal axis and $L$ on the vertical one.
Red lines are creation resonances, $(\omega_i+\omega_j)/2\pi$, and black
lines are exchange resonances, $|\omega_i-\omega_j|/2\pi$. Each point where two lines cross gives a value of $L$ at which the frequency drives two processes. Purple triangles mark crossings of two creation lines which involve four modes. Yellow squares and gray circles mark
crossings of a creation line with an exchange line. If both processes share a mode and creation coupling is stronger than exchange coupling, the crossing gives a three mode resonance (yellow squares). Otherwise (gray circles), either the two processes do not share a mode and run
independently, or exchange is stronger and stops the growth. Gray circles
fill the low-frequency part of the map, while yellow squares appear only
for $\Omega/2\pi\gtrsim17$GHz.

{\textbf{Three-mode resonance.}}
At a mixed crossing the pump satisfies $\Omega\approx\omega_k+\omega_l$ (creation) and $\Omega\approx\omega_m-\omega_k$ (exchange) at the same time, with mode $k$ shared by both processes. The RWA now retains two sets of terms: the pair-creation term $\hat a_k^\dagger\hat a_l^\dagger e^{-i\Omega t}$ and the exchange term $\hat a_m^\dagger\hat a_k e^{-i\Omega t}$, together with their Hermitian conjugates, with strengths $g_{kl}$ and $g_{km}$ by the same counting as in the two-mode case. The rotating frame must remove both phases, which requires $\omega_{{\rm rot},k}+\omega_{{\rm rot},l}=\Omega$ and $\omega_{{\rm rot},m}-\omega_{{\rm rot},k}=\Omega$. These are two conditions for three frequencies, so one choice remains free; we take $\omega_{{\rm rot},k}=\omega_k$, $\omega_{{\rm rot},l}=\Omega-\omega_k$ and $\omega_{{\rm rot},m}=\Omega+\omega_k$, which places the detunings $\Delta_1=\Omega-(\omega_k+\omega_l)$ and $\Delta_2=\Omega-(\omega_m-\omega_k)$ on modes $l$ and $m$, respectively. Any other choice only redistributes the detunings among the modes and leaves $\langle\hat N_i\rangle$ unchanged (Appendix~\ref{app:rwa_and_rot}). The effective Hamiltonian reads

\begin{eqnarray}
    \hat{H}_{\text{eff}}/\hbar = -\Delta_1 \hat{a}_l^\dagger \hat{a}_l - \Delta_2 \hat{a}_m^\dagger \hat{a}_m + g_{kl}(\hat{a}_k^\dagger \hat{a}_l^\dagger + \hat{a}_k \hat{a}_l) \nonumber\\+ g_{km}(\hat{a}_m^\dagger \hat{a}_k + \hat{a}_m \hat{a}_k^\dagger)
    \label{eq:Hthreemode}
\end{eqnarray}

The two processes compete for mode $k$: the $g_{kl}$ term creates pairs while the $g_{km}$ term moves particles from $k$ to $m$. At exact resonance ($\Delta_1=\Delta_2=0$), the Heisenberg equations close on the set $(\hat a_k,\hat a_l^\dagger,\hat a_m)$,
\begin{equation}
\begin{aligned}
\dot{\hat a}_k &= -ig_{kl}\hat a_l^\dagger-ig_{km}\hat a_m,\\
\dot{\hat a}_l^\dagger &= ig_{kl}\hat a_k,\qquad
\dot{\hat a}_m = -ig_{km}\hat a_k,
\end{aligned}
\end{equation}
which give $\ddot{\hat a}_k=\Gamma^2\hat a_k$ with $\Gamma = \sqrt{g_{kl}^2 - g_{km}^2}$. Integrating them from the vacuum,
\begin{equation}
\begin{aligned}
\langle \hat N_k(t)\rangle &= \frac{g_{kl}^2}{\Gamma^2}\sinh^2(\Gamma t), \\ 
\langle \hat N_m(t)\rangle &= \frac{g_{kl}^2 g_{km}^2}{\Gamma^4}\left[\cosh(\Gamma t)-1\right]^2,  \\
\langle \hat N_l(t)\rangle &= \frac{g_{kl}^2}{\Gamma^2}\sinh^2(\Gamma t)
+\frac{g_{kl}^2 g_{km}^2}{\Gamma^4}\left[\cosh(\Gamma t)-1\right]^2  \\ &= \langle \hat N_k(t)\rangle+\langle \hat N_m(t)\rangle. \label{eq:N3mode}
\end{aligned}
\end{equation}
Each created pair puts one particle in $l$ and one in $k$, and exchange then moves part of the latter to $m$, so $\hat N_l-\hat N_k-\hat N_m$ is conserved. Populations grow exponentially only if $\Gamma$ is real, that is, if 
 $|g_{kl}|>|g_{km}|$.
Since $g_{kl}^2=g_{kk}g_{ll}$ and $g_{km}^2=g_{kk}g_{mm}$ (Eq.~\eqref{eq:16}), this condition becomes  $|g_{ll}|>|g_{mm}|$:
mode $k$ drops out, and creation wins only if mode $l$ couples to the SQUID more strongly
than mode $m$.

\begin{figure}
    \centering
    \includegraphics[width=0.85\linewidth]{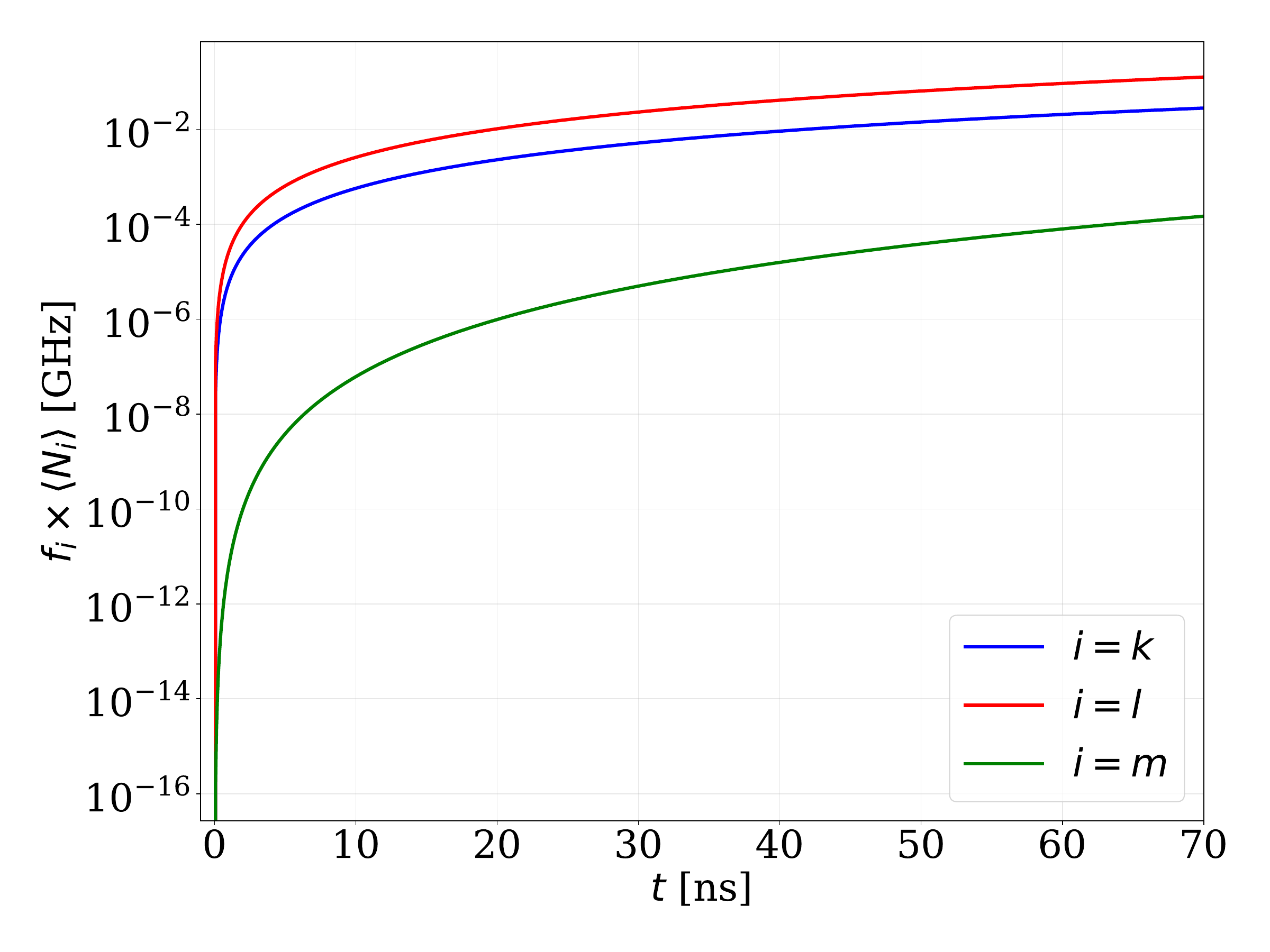}
    \caption{Time evolution of the modal energy $f_i \times \langle N_i \rangle$ ($f_i=\omega_i/2\pi$) during a three-mode resonance resulting from a mixed creation-exchange degeneracy. 
    The target mode frequencies are $f_k = 3.6$GHz, $f_l = 16.3$GHz, and $f_m = 23.6$GHz, with effective couplings $g_{kl}/2\pi = 0.2$MHz and $g_{km}/2\pi = 0.1$MHz. Parameters are identical to those in Fig. \ref{fig:barridoindL}, choosing one specific yellow crossing mark.}
    \label{fig:threemode}
\end{figure}

Fig.~\ref{fig:threemode} shows the energy $f_i\langle N_i\rangle$ of the three
modes at a yellow crossing with $f_k=3.6$GHz, $f_l=16.3$GHz and $f_m=23.6$GHz,
where $g_{kl}>g_{km}$. Since $g_{kl}/g_{km} = (g_{ll}/g_{mm})^{1/2}$, this requires $g_{ll} > g_{mm}$ although $f_l < f_m$: with these parameters the LH coupling does not keep growing up to the highest modes.
All three modes are populated. Modes $k$ and $l$ grow from the start, as in a two-mode resonance, while mode $m$ is filled only through $k$ and stays behind. For short times 
($\Gamma t\ll1$), Eq.~\eqref{eq:N3mode} gives
$\langle N_k\rangle\approx g_{kl}^2t^2$ and
$\langle N_m\rangle\approx g_{kl}^2g_{km}^2t^4/4$.
The exchange
also lowers the growth rate, $\Gamma<g_{kl}$. A mixed crossing therefore
adds a third mode, at the cost of slower growth.

{\textbf{Four-mode resonance.}}
At the crossing of two creation lines,  $\Omega \approx \omega_k + \omega_l \approx \omega_m + \omega_n$. The two pairs cannot share a mode: $k=m$ would give $\omega_l=\omega_n$, and hence $l=n$, since the spectrum is non-degenerate. 

The RWA retains the two pair-creation terms, $g_{kl}\hat a_k^\dagger\hat a_l^\dagger e^{-i\Omega t}$ and $g_{mn}\hat a_m^\dagger\hat a_n^\dagger e^{-i\Omega t}$, plus their Hermitian conjugates. Cross terms between the two pairs, such as $\hat a_k^\dagger\hat a_m^\dagger$, oscillate at $\omega_k+\omega_m-\Omega=\omega_m-\omega_l$, of the order of the mode spacing, and are dropped. The rotating frame now has four frequencies and two conditions, $\omega_{{\rm rot},k}+\omega_{{\rm rot},l}=\Omega$ and $\omega_{{\rm rot},m}+\omega_{{\rm rot},n}=\Omega$. Defining the independent detunings $\Delta_1 = \Omega - (\omega_k + \omega_l)$ and $\Delta_2 = \Omega - (\omega_m + \omega_n)$, we split each detuning symmetrically within its pair, as in the two-mode case. The effective Hamiltonian is then the sum of two commuting terms of the form of Eq.~\eqref{eq:Htwomode}, one per pair, and the Heisenberg equations decouple into two independent two-mode systems. Each pair evolves as in Eq.~\eqref{eq:ak(t)} and  Eq.~\eqref{eq:al(t)}, with $\Gamma_{kl}=\sqrt{g_{kl}^2-(\Delta_1/2)^2}$ and $\Gamma_{mn}=\sqrt{g_{mn}^2-(\Delta_2/2)^2}$.

\begin{figure}
    \centering
    \includegraphics[width=0.85\linewidth]{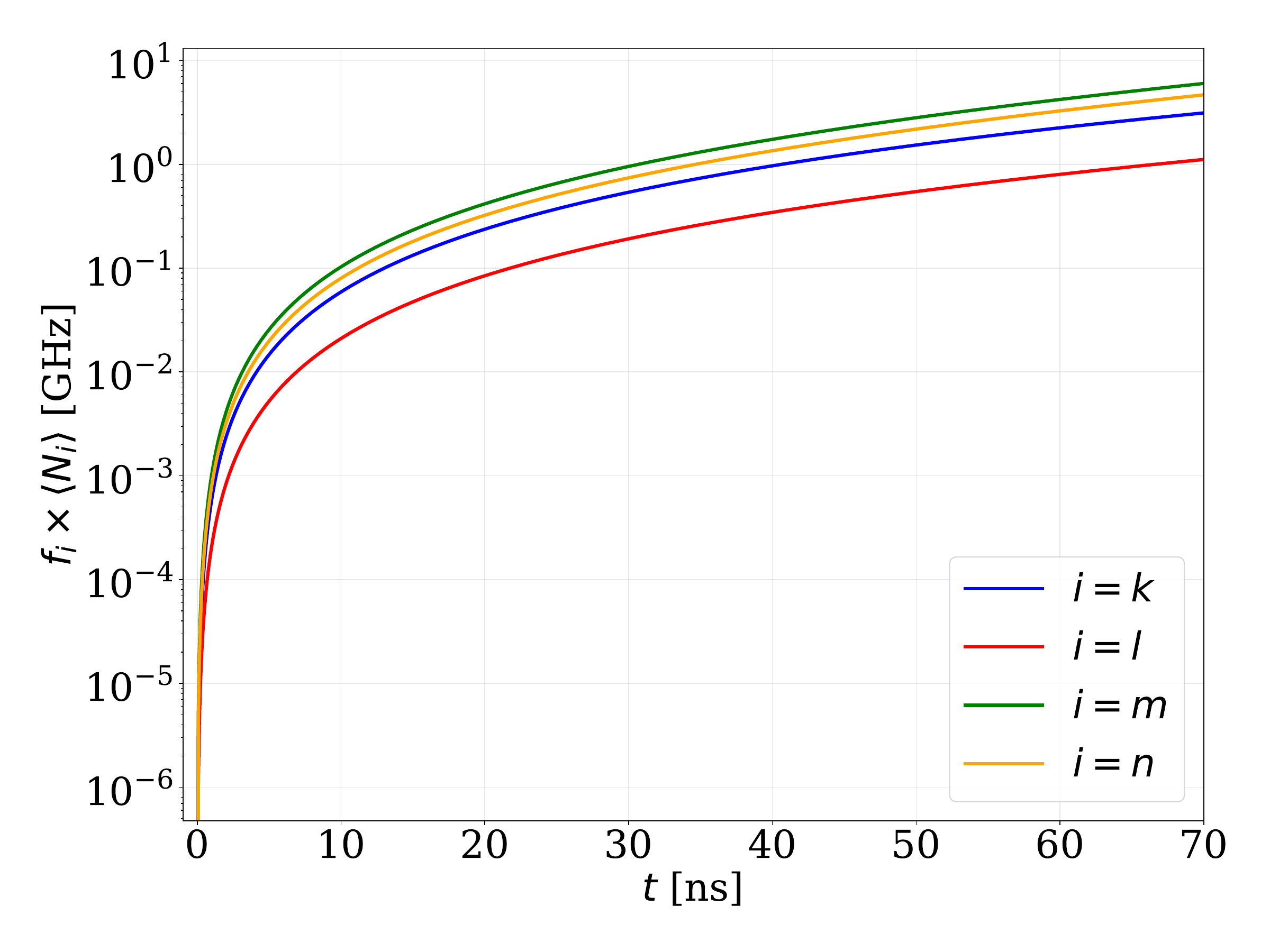}
    \caption{Time evolution of the modal energy $f_i \times \langle N_i \rangle$ ($f_i=\omega_i/2\pi$) during a four-mode resonance resulting from a two-pair creation degeneracy.
    Participating mode frequencies are $f_k = 12.1$GHz, $f_l = 4.3$GHz, $f_m = 9.2$GHz, and $f_n = 7.2$GHz, driven by couplings $g_{kl}/2\pi = 1.1$MHz and $g_{mn}/2\pi = 1.7$MHz. Parameters are identical to those in Fig. \ref{fig:barridoindL}, choosing one specific purple crossing mark.}
    \label{fig:fourmode}
\end{figure}

Fig.~\ref{fig:fourmode} shows the four modes at a purple crossing. Within each pair $\langle N_k \rangle= \langle N_l \rangle$ and $\langle N_m \rangle= \langle N_n \rangle$, so the two curves of a pair differ only by the factor $f_k/f_l$ or $f_m/f_n$. 
For $\Gamma t \ll 1$, Eq.~\eqref{eq:NkNl}  gives
$\langle N_k \rangle= \langle N_l \rangle \approx g_{kl}^2 t^2$ and $\langle N_m \rangle= \langle N_n \rangle \approx g_{mn}^2 t^2$, so 
the pair $(m,n)$ is more populated by a factor $(g_{mn}/g_{kl})^2$.
Crossings therefore populate more modes, but do not increase the growth rate: two creation processes run independently, and an exchange process slows creation down.

\section{Single resonance in dense spectrum regions}
\label{sec:densespectrum}

In Fig. \ref{fig:barridoindL} the resonance lines become dense at low $\Omega$, where both modes lie near the lower band edge. 
One might wonder if the  single-resonance approximation of Sec. III could fail, since it neglects nearby pairs. However, near the band edge the coupling is small (Fig.~\ref{fig:amplificacionyatenuacion}). Because the coupling for these modes is much smaller than the remaining frequency gap, the single-resonance approximation holds, except at the exact degeneracy points (crossings) of Sec.~\ref{sec:multipleresonance}.

Increasing $N$ makes the spectrum near the band edge denser. Numerically, for $\eta \ll 1$, we find no excitation of many modes in this region. 

To understand why this multimode explosion does not occur, we analyze the asymptotic scaling of both the frequency spacing $\Delta\omega$ and the coupling strength $g$ near the edge of the Brillouin zone as the number of unit cells $N$ increases. Using the dimensionless wavevector and expanding the transcendental equation for small deviations from the zone boundary, we find that the frequency spacing between adjacent modes scales as
\begin{equation}
    \Delta\omega = \omega_i - \omega_{i+1} \sim \mathcal{O}\left(\frac{1}{N^2}\right) \text{.}
\end{equation}

However, evaluating the exact  coupling in this same asymptotic limit reveals that the oscillatory terms and the normalization lead to a much steeper asymptotic decay for the coupling constant:
\begin{equation}
    g \sim \mathcal{O}\left(\frac{1}{N^3}\right) \text{.}
\end{equation}

\begin{figure}[t]
    \centering\includegraphics[width=0.85\linewidth]{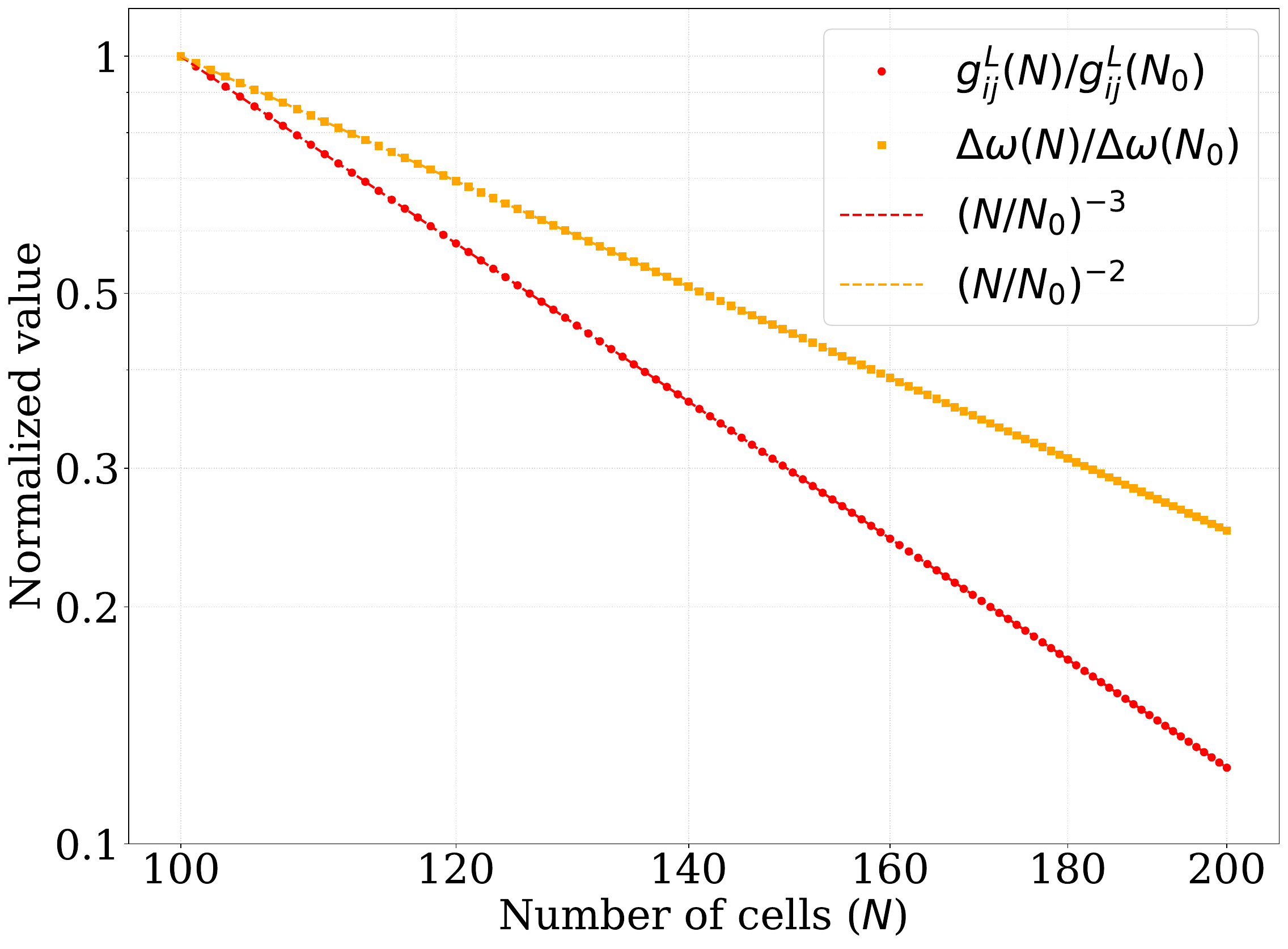}
    \caption{Coupling $g^L_{ij}$ and frequency spacing $\Delta\omega$ between the two modes closest to the LH band edge as a function of the number of cells $N$, each normalized to its value at $N_0 =100$ cells, where $g^L_{ij}(N_0)/2\pi = 3\times10^{-7}$MHz and $\Delta\omega(N_0)/2\pi = 0.93$MHz. Dashed lines show $(N/N_0)^{-3}$ and $(N/N_0)^{-2}$. The coupling decays faster than the spacing ($N^{-3}$ versus $N^{-2}$) and stays about $6.6$ orders of magnitude below it, so increasing the number of cells does not lead to multimode excitation. Other parameters as in Fig.~\ref{fig:frecuencias}.}
    \label{fig:gvsD}
\end{figure}

This analytical scaling is consistent with the numerical results of Fig. \ref{fig:gvsD}: the coupling strength decays much faster than the frequency spacing, and the ratio $g/\Delta\omega$ is already of order $3.2\times10^{-7}$ at $N = 100$. Therefore, as we increase $N$ to maximize the modal density, the inter-mode couplings are suppressed significantly faster than the frequency gaps. This explains why the intuitive strategy of enhancing the mode density by extending the length of the transmission line does not lead to simultaneous multimode excitation.

\section{LHST versus RHST and CPW performance}
\label{sec:LvsRperfo}

So far we have compared the two lines at equal circuit elements and for a few selected modes ($N=30$). Here we ask which line is preferable for a given pump frequency, and how the LHST compares with a coplanar waveguide (CPW) terminated by the same SQUID, the geometry used in DCE experiments~\cite{Wilson2010, Wilson2011} and studied theoretically in Refs.~\cite{Wustmann2013, Lombardo2016}.

\begin{figure*}
\centering
\includegraphics[width=0.75\linewidth]{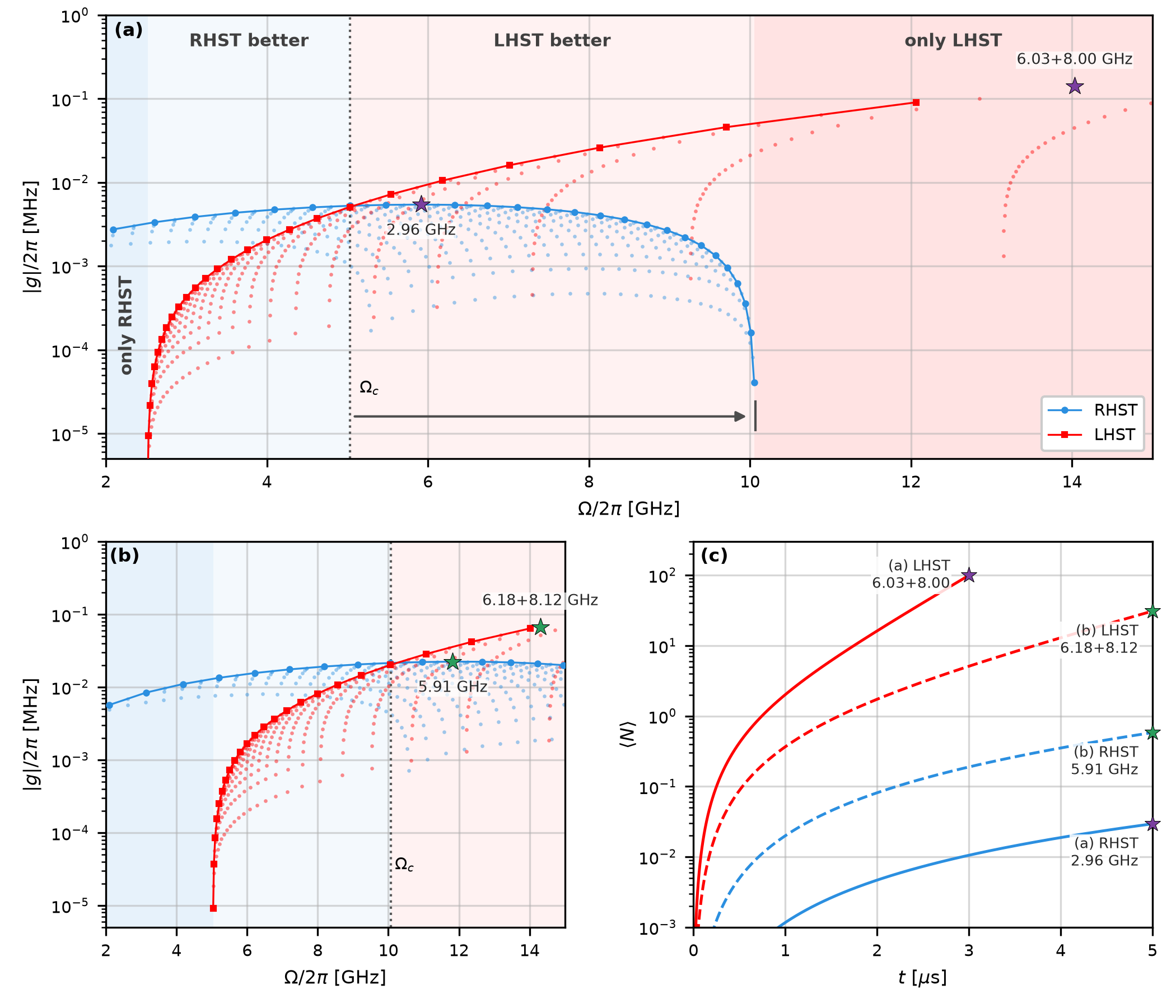}
\caption{Parametric couplings of the LHST (red squares) and RHST (blue circles) as a function of the pump frequency $\Omega$, for (a) $L=2$nH, $C=2$pF and (b) $L=1$nH, $C=1$pF,  the same parameters as in Fig.~\ref{fig:amplificacionyatenuacion}. Large markers joined by lines are single-mode resonances, $\Omega=2\omega_k$, with coupling $g_{kk}$; small dots are two-mode resonances, $\Omega=\omega_k+\omega_l$, with coupling $g_{kl}$. Shaded regions mark where only one line has resonances  and, where both resonate, which line has the larger single-mode coupling.  The shading in (b) follows the same convention. The arrow in (a) shows the shift of the crossing $\Omega_c$ from $L=2$nH, $C=2$pF  to $L=1$nH, $C=1$pF. Stars mark the strongest resonance of each line in the window 2--15~GHz, and the labels give the frequencies of the modes involved. For the LHST this is a two-mode resonance between neighboring modes, so its star lies among the small dots. (c) Number of particles created by the starred resonances, $\langle N\rangle=\sinh^2(g_{kk}t)$ for a single mode and $\langle N_k+N_l\rangle=2\sinh^2(g_{kl}t)$ for a pair; solid (dashed) lines correspond to the stars in panel (a) [(b)]. Other parameters: $N=30$, $I_J=10\mu$A, $C_J=0.1$pF, $\eta=0.01$.}
    \label{ventana}
\end{figure*}

In an experiment the circuit is fixed and the pump frequency is the parameter one can tune, within the range accessible to flux pumping \cite{Wilson2010, Wilson2011, Schneider2020}. Figure~\ref{ventana} collects the single- and two-mode resonances of Secs.~\ref{1res} and ~\ref{2res} as a function of the pump frequency $\Omega$, within a window of 2–15GHz for the RHST and LHST. At low pump frequencies only the RHST has resonances, since the LHST spectrum starts at $\omega_{\min}$. At high pump frequencies only the LHST has them, since the RHST spectrum ends at its band edge. In between, the two lines cross at $\Omega_c\simeq 2/\sqrt{LC}$. This follows from the dispersion relations, which satisfy $\omega({\chi})\nu({\chi})=1/LC$ at equal wave vector $\chi$, so both lines have the same wave vector and similar couplings at $\omega=\nu=1/\sqrt{LC}$. Since the number of created particles grows as $\sinh^2(gt)$, the line with the larger coupling creates particles faster: below $\Omega_c$ the RHST is preferable, at $\Omega_c$ both are equivalent, and above $\Omega_c$ the LHST is preferable; this holds for equal circuit elements. Larger $L$ and $C$ lower $\Omega_c$ and the RHST band edge together, so the LHST covers a wider range of pump frequencies (comparison between panel (a) and (b)). 

The stars in Fig.~\ref{ventana} mark the strongest resonance of each line within the pump window. For the RHST it is a single-mode resonance near the maximum of its coupling. For the LHST it is a pair of neighboring modes, by the same mechanism as in Fig.~\ref{fig:ampl_twomode}: the pair coupling borrows from the higher, more strongly coupled mode, and the pump window limits the partner to the neighboring mode. Going from (b) to (a), the LHST resonance becomes stronger, while the best RHST resonance becomes weaker because its band shrinks. Panel (c) shows the particles created by these resonances. Because the growth is exponential in the coupling, a moderate difference in $g$ becomes a large difference in particle number, and larger $L$ and $C$ widen the gap between the two lines.

The same inductors, capacitors and SQUID can be arranged either as a conventional  array (RHST) or as a left-handed one (LHST). Which arrangement is preferable depends on the frequency at which the external flux can be modulated. If the available pump lies below $\Omega_c\simeq 2/\sqrt{LC}$, the conventional array creates particles faster. If it lies above $\Omega_c$, the left-handed array does, and above the RHST band edge it is the only option. Since $\Omega_c$ depends only on the cell elements, it can be placed where needed: larger $L$ and $C$ bring it down, so that the left-handed array becomes preferable for the pump frequencies reached in flux-pumping experiments.

\begin{figure*}
\centering
\includegraphics[width=0.65\linewidth]{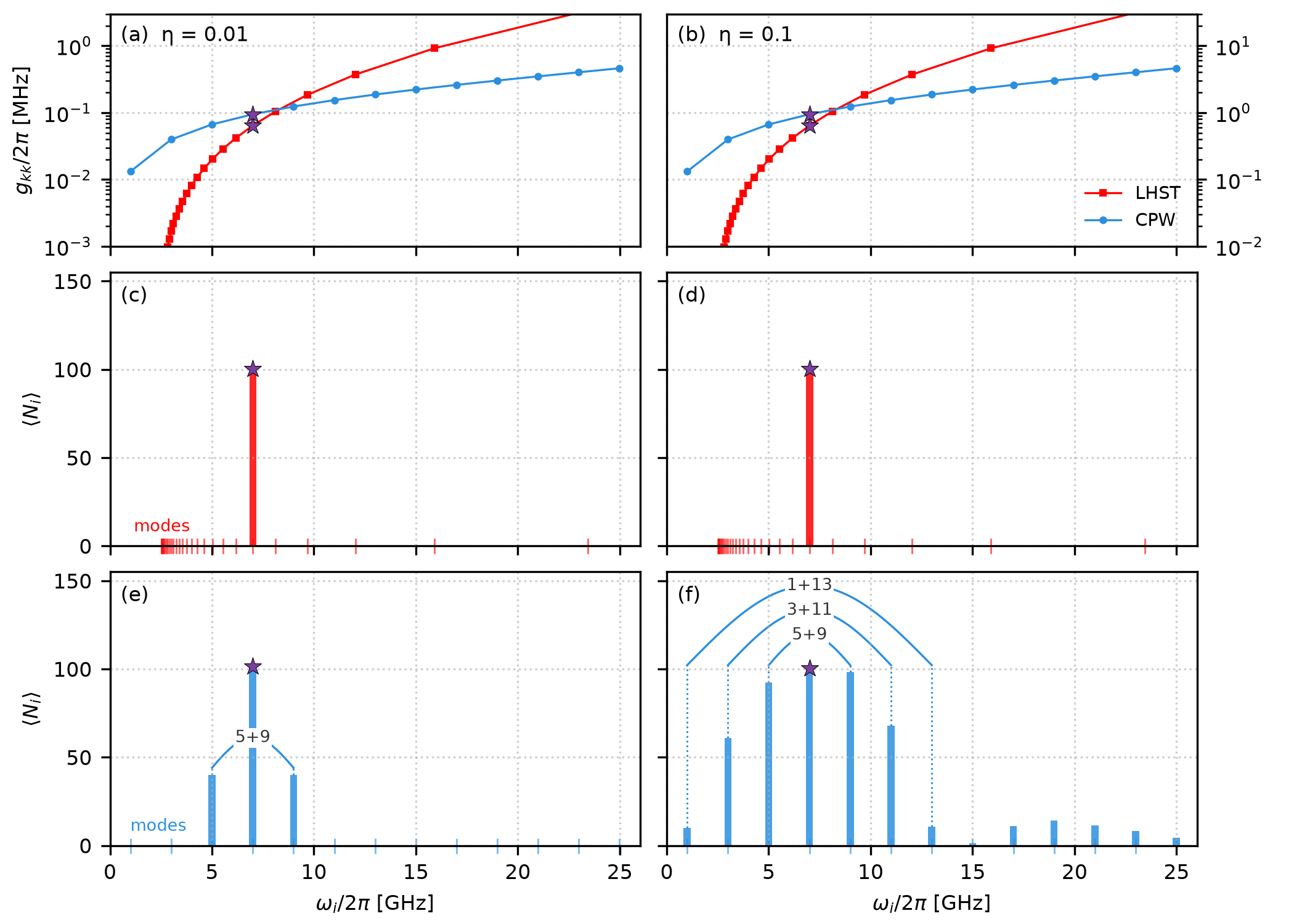}
    \caption{Single-pair versus multi-pair creation in the LHST (red) and in a long coplanar waveguide (CPW, blue) terminated by the same SQUID. Both are pumped at $\Omega/2\pi=14$GHz, twice the frequency of a target mode at 7GHz (star). (a), (b) Single-mode couplings $g_{kk}$ of all modes for $\eta=0.01$ and $\eta=0.1$, respectively. At the target, $g_{kk}/2\pi=0.064$MHz (LHST) and $0.094$MHz (CPW) for $\eta=0.01$, and ten times larger for $\eta=0.1$. (c)--(f) Mean number of particles in each mode when the target has reached $\langle N_k\rangle \approx \sinh^2 3 \approx 100$, the population of an isolated resonance at $g_{kk}t=3$: (c), (d) LHST and (e), (f) CPW, for $\eta=0.01$ (left) and $\eta=0.1$ (right). Ticks on the horizontal axis mark the positions of all modes. Arcs join the two modes of each pair that satisfies $\omega_i+\omega_j\simeq\Omega$, labeled by their frequencies in GHz. The energy cost, defined as the total created energy divided by the energy in the target, $\sum_i\omega_i\langle N_i\rangle/(\omega_k\langle N_k\rangle)$, is 1.0 for the LHST at both $\eta$, and 1.8 ($\eta=0.01$) and 6.0 ($\eta=0.1$) for the CPW, which for $\eta=0.1$ needs $g_{kk}t\approx 4$ to reach this population. LHST: $N=30$, $L=1$nH, $C=1$pF. CPW: $Z_0=50\Omega$, $v=1.2\times10^8$m/s, far end open and fundamental mode at 1GHz ($L_{\rm tot}/L_J\approx380$). In both lines, $I_J=10\mu$A and $C_J=0.1$pF.}
    \label{aislamiento}
\end{figure*}
We now compare the LHST with a long CPW cavity terminated by the same SQUID, pumped at twice the frequency of a target mode (Fig.~\ref{aislamiento}). 
The CPW is a continuous RH line, 
so its modes are nearly equidistant, and with $L_{\rm tot}/L_J \approx 380$, changes this only slightly \cite{Lombardo2016}. The LHST spectrum, instead, becomes sparse and far from equidistant at high frequencies (Sec.\ref{sec:model}). 
Panels (a) and (b) show the couplings for $\eta=0.01$ and $\eta=0.1$, respectively. By Eq.~\eqref{eq:16}, $g_{ij}\propto\alpha\propto\eta$ in both lines, so the curves in (b) are those of (a) multiplied by 10. At the target the two lines have similar couplings, so an isolated target would grow at a similar rate in both. Panels (c)--(f) show how the created particles are distributed among the modes when the target has reached the same population. In the LHST [(c) and (d)] all particles are in the target mode, for weak and strong pumping. In the CPW [(e) and (f)], other modes are also populated, and more of them as the pumping increases.

The difference comes from the spectrum. The modes of a long CPW are nearly equidistant, so every pair placed symmetrically around the target also satisfies $\omega_i+\omega_j\simeq\Omega$. Each such pair is driven at the same time as the target, as in the creation-creation crossings of four-mode resonance, and it grows whenever the threshold condition of Eq.~\eqref{resonancecond} holds. Exchange processes with $|\omega_i-\omega_j|\simeq\Omega$ are resonant for the same reason and move particles between modes. The couplings grow with $\eta$, while the detunings are fixed by the spectrum. Stronger pumping therefore brings more pairs into resonance, which is why the number of populated modes increases from (e) to (f). The LHST spectrum is far from equidistant, so the detuning of the nearest competing pair is much larger than its coupling, even for strong pumping. Its many modes near the band edge, visible as ticks in (c) and (d), remain empty.

Pairs that do not share a mode with the target evolve independently and do not slow its growth. They still take energy from the pump and leave the line in a multimode state, which adds noise to a measurement of the target. Exchange processes that share the target mode act as in the three-mode crossings of Sec.~\ref{sec:multipleresonance}: they populate additional modes and slow the growth of the target. In the nearly equidistant CPW every populated mode has a partner $\Omega$ above it, so exchange also populates the modes above the target, as seen in (f). 

To reach the same population of the target, the CPW therefore spends more pump energy than the LHST. Each created pair takes one quantum $\hbar\Omega$ from the pump, wherever it ends up. In the LHST all of this energy goes into the target, whereas in the CPW part of it goes into the other pairs and their exchange partners. The energy cost defined in the caption of Fig.~\ref{aislamiento} grows with the pumping: the CPW needs about twice the energy of the LHST at weak pumping and about six times at strong pumping. The CPW reaches this population somewhat earlier because its coupling is slightly larger, but the advantage of the LHST lies in where the energy goes. Its non-equidistant spectrum removes the systematic coincidences of an equidistant line, but not accidental ones, which may occur for particular choices of $L$, $C$ and pump frequency.

A CPW cavity can be made less equidistant with a SQUID that carries a larger fraction of the inductance, i.e, a smaller $L_{\rm tot}/L_J $ \cite{Wustmann2013}. The departure from equidistance remains small, however, so the  pumped mode stays isolated only while its exchange coupling to the next mode is smaller than this departure.

\section{Conclusion}
\label{sec:conclusion}

We have studied the dynamical Casimir effect in a left-handed transmission line terminated by a flux-modulated SQUID, and compared it with a right-handed line built from the same elements and with a long conventional coplanar waveguide terminated by the same SQUID.
We have derived the quantum Hamiltonian of both lines. In both cases, the coupling 
 factorizes as in Eq.\eqref{eq:16}, so the differences come from the spectrum and from the mode amplitudes at the SQUID. The LH line has a lower band edge and extends to high frequencies, where its modes are sparse, whereas the RH line has an upper band edge.

Under single-mode pumping, the LH coupling increases with frequency, while the RH coupling remains nearly constant except near its upper band edge. At equal mode frequencies, the LH line therefore produces fewer particles near its lower band edge
and more at high frequencies, and only the LH line has modes above the RH band edge, where each created particle carries more energy. 

Under two-mode pumping, the factorized coupling enables a weakly coupled low-frequency LH mode to be paired with a strongly coupled high-frequency mode. The population of the target mode can then exceed the best RH option, at the cost of an equal number of particles in the partner mode.  

Since the inductance $L$ enters the mode equation through $L/L_J$, the resonance lines shift non-uniformly with $L$ and cross in the $(\Omega, L)$ plane, where a single pump frequency drives two processes. When a creation resonance intersects an exchange resonance sharing a mode, a third mode becomes populated and the growth rate is reduced; at a creation-creation crossing, the two resonant pairs evolve independently. Crossings therefore populate more modes but do not increase the growth rate.  Near the LH band edge the couplings decrease as $N^{-3}$ while the mode spacing decreases as $N^{-2}$, so longer lines do not lead to multimode excitation.

The comparison between the two arrays can be read as a design rule: given a set of inductors, capacitors and a SQUID, whether to arrange them as a conventional or as a left-handed array depends on the range over which the external flux can be modulated and on the purpose. The conventional array creates particles faster at low pump frequencies, the left-handed one at high pump frequencies, and only the left-handed array has resonances above the band edge of the conventional one. The crossover, $\Omega_c \simeq 2/\sqrt{LC}$, is set by the cell inductance and capacitance; larger values lower it, so the left-handed array becomes preferable over a wider range of pump frequencies. At very high frequencies, however, parasitic RH elements not included in our model may become relevant~\cite{LaiItohCaloz2004,Caloz2004,CalozItohBook}.

Compared with the long coplanar waveguide, the LH line has a similar coupling but a very different spectrum. The nearly equidistant waveguide drives many pairs together with the target, so it needs more pump energy to reach the same target population, and this cost grows with the modulation depth. The LH spectrum keeps the created particles in the target mode, although accidental coincidences may occur for particular parameters. A waveguide in which the SQUID carries a larger fraction of the inductance is less equidistant, but keeps the target isolated only while the pumping is weak.

Beyond particle creation in selected modes, LH lines may also serve as quantum simulators of fields with unconventional dispersion relations, such as massive fields with left-handed group velocity~\cite{FerreriWilhemCreation}.

\section*{Acknowledgments}

 This work was supported by Consejo Nacional de Investigaciones Científicas y Técnicas (CONICET) and Universidad de Buenos Aires (UBA).

\appendix

\section{Equations of Motion and Matrix Formalism}
\label{app:eqs_motion}

In Sec.~\ref{sec:model}, the Lagrangian of the transmission lines was constructed by identifying the kinetic and potential energies associated with the capacitive and inductive elements, respectively. Alternatively, one can derive this Lagrangian directly from the classical equations of motion, which enforce current conservation (Kirchhoff's current laws) at each node of the circuit~\cite{YurkeDenker1984, VoolDevoret2017}. We present this derivation here in parallel for both the LHST and RHST arrays.

For the bulk nodes ($n=1,\dots,N-1$), the topological arrangement of the components dictates the flow of currents. In the LHST line, series capacitors and shunt inductors lead to the following dynamical equations for the node fluxes:
\begin{equation}
\frac{d}{dt}\Big[-C(\dot\phi_{n+1}-\dot\phi_n)+C(\dot\phi_n-\dot\phi_{n-1})\Big] = -\frac{\phi_n}{L}.
\label{eq:bulk_LH}
\end{equation}
In contrast, exchanging the series and shunt elements for the RHST line yields a dual nodal equation:
\begin{equation}
\frac{d}{dt}\Big[C\dot\phi_n\Big] = \frac{\phi_{n+1}-\phi_n}{L} - \frac{\phi_n-\phi_{n-1}}{L}.
\label{eq:bulk_RH}
\end{equation}
Both configurations assume a grounded boundary condition $\phi_0=0$ at the left end of the array.

At the right boundary ($n=N$), the presence of the SQUID modifies the current conservation equations by introducing its intrinsic capacitance $C_J$ and a nonlinear Josephson current $I_s = 2I_J\cos\big(\frac{\Phi_{\text{ext}}(t)}{2\varphi_0}\big)\sin\big(\frac{\phi_N}{\varphi_0}\big)$ (with $\varphi_0=\frac{\Phi_0}{2\pi}$), modulated by the external flux $\Phi_{\text{ext}}(t)$~\cite{apC1, apC2, Wallquist2006}. The equations of motion at the boundary for the LHST and RHST lines are, respectively:
\begin{align}
\text{LHST:} \quad &\frac{d}{dt}\Big[C(\dot\phi_N-\dot\phi_{N-1})+2C_J\dot\phi_N\Big] = -I_s - \frac{\phi_N}{L}, \label{eq:bound_LH} \\
\text{RHST:} \quad &\frac{d}{dt}\Big[(C+2C_J)\dot\phi_N\Big] = -\frac{\phi_N-\phi_{N-1}}{L} - I_s. \label{eq:bound_RH}
\end{align}

Assuming the system operates in the linear regime ($\phi_N \ll \varphi_0$), we approximate $\sin(\phi_N/\varphi_0) \approx \phi_N/\varphi_0$. The parametrically driven Josephson energy is defined as $E_J(t) = 2I_J\varphi_0 \cos\big(\frac{\Phi_{\text{ext}}(t)}{2\varphi_0}\big) \equiv E_0[1+\eta\cos(\Omega t)]$. Using the static Josephson inductance relation $1/L_J \equiv E_0/\varphi_0^2$ and assuming a weak driving amplitude $\eta \ll 1$, the Josephson current linearizes to $I_s \approx \frac{\phi_N}{L_J} + \frac{\eta\cos(\Omega t)}{L_J}\phi_N$. 

Substituting this linearized current back into Eqs.~\eqref{eq:bound_LH} and \eqref{eq:bound_RH}, the full sets of coupled linear differential equations for both arrays correspond exactly to the Euler-Lagrange equations of the compact matrix Lagrangian presented in Eq.~\eqref{eq:4}:
\begin{equation}
\mathcal L = \frac{C}{2}\dot{\bar\phi}^T\mathbb A\,\dot{\bar\phi}
-\frac{1}{2L}\bar\phi^T\mathbb B\,\bar\phi
-\frac{\eta\cos(\Omega t)}{2L_J}\phi_N^2.
\end{equation}

The structural differences between the two transmission lines are fully absorbed into the dimensionless capacitance ($\mathbb A$) and inductance ($\mathbb B$) matrices. 

For the LH line, the series capacitors couple adjacent nodes dynamically, resulting in a tridiagonal matrix $\mathbb A_{LH}$, while the shunt inductors connect each node to ground, making $\mathbb B_{LH}$ diagonal:
\begin{align}
\mathbb A_{LH} &= \begin{pmatrix}2&-1&0&\dots&0\\-1&2&-1&\dots&0\\0&-1&2&\dots&0\\
\vdots&\vdots&\vdots&\ddots&-1\\0&0&\dots&-1&1+\frac{2C_J}{C}\end{pmatrix},
\label{eq:A-LH}\\
\mathbb B_{LH} &= \mathrm{diag}\left(1,1,\dots,1,\,1+\frac{L}{L_J}\right).
\label{eq:B-LH}
\end{align}

Conversely, for the RH line, the topological inversion dictates that capacitors shunt to ground and inductors couple in series. This exactly reverses the structures, yielding a diagonal $\mathbb A_{RH}$ and a tridiagonal $\mathbb B_{RH}$:
\begin{align}
\mathbb A_{RH} &= \mathrm{diag}\left(1,1,\dots,1,\,1+\frac{2C_J}{C}\right), \label{eq:A-RH}\\
\mathbb B_{RH} &= \begin{pmatrix}2&-1&0&\dots&0\\-1&2&-1&\dots&0\\0&-1&2&\dots&0\\
\vdots&\vdots&\vdots&\ddots&-1\\0&0&\dots&-1&1+\frac{L}{L_J}\end{pmatrix}. \label{eq:B-RH}
\end{align}

The diagonalization of these matrices and the ensuing canonical quantization steps follow identically for both systems, leading directly to the normal modes and couplings summarized in Table~\ref{tab:summary}.

\section{Robustness of Coupling Ranges Across Junction Parameters}
\label{app:couplings}

\begin{figure}[htbp]
    \centering
    \begin{subfigure}{0.45\columnwidth}
        \centering
        \includegraphics[width=\linewidth]{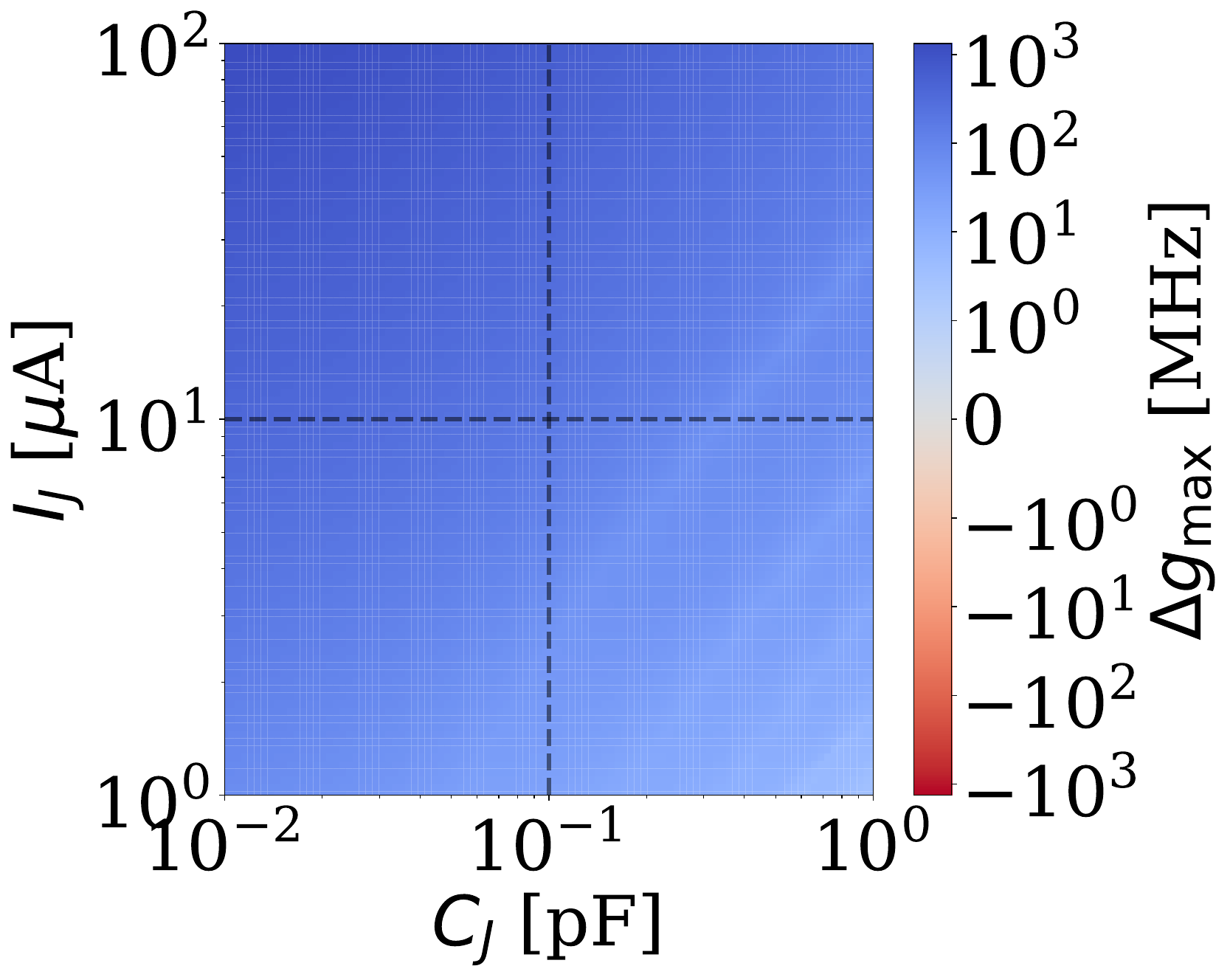}
        \label{fig:subfig1_app}
    \end{subfigure}
    \begin{subfigure}{0.45\columnwidth}
        \centering
        \includegraphics[width=\linewidth]{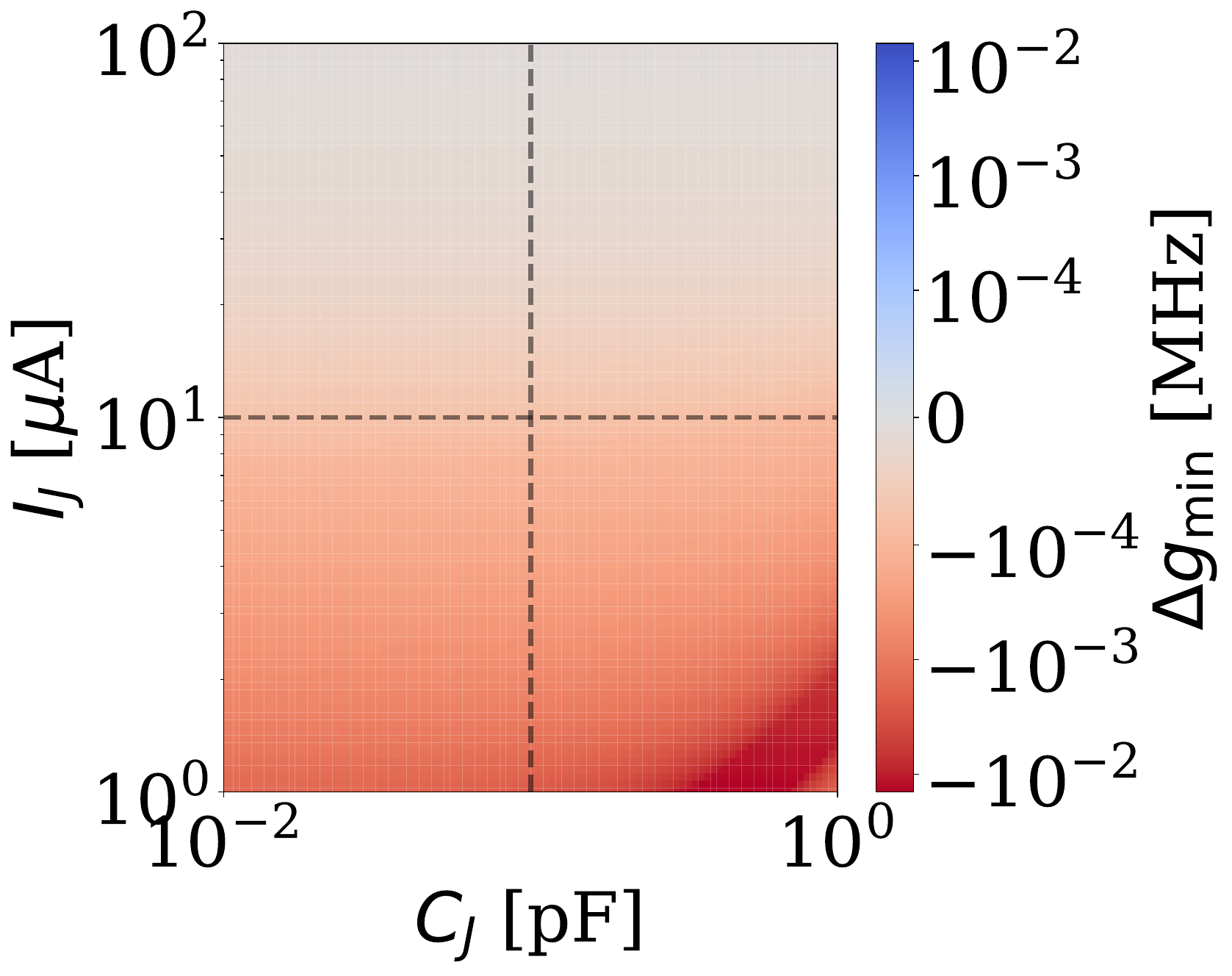}
        \label{fig:subfig2_app}
    \end{subfigure}
    \caption{Heatmaps of the maximum ($\Delta g_{\max}=g_{\max}^L-g_{\max}^R$, left) and minimum ($\Delta g_{\min}=g_{\min}^L-g_{\min}^R$, right) single-mode coupling differences between the LHST and RHST architectures as a function of the junction parameter space $(C_J, I_J)$. The parameter ranges encompass values reported for standard superconducting junctions~\cite{apC1,apC2,apC3,apC4,apC5,apC6}. The horizontal and vertical dashed lines mark the nominal circuit parameters $I_J = 10\mu\text{A}$ and $C_J = 0.1\text{pF}$ used in the main text.}
    \label{fig:barridoC_JyI_J}
\end{figure}

Since the allowed wavevectors $k_i$ and $p_i$ are roots of the transcendental boundary equations, the resulting coupling strengths $g_{ii}^{L,R}$ depend implicitly on the SQUID junction capacitance $C_J$ and critical current $I_J$ (which sets the static inductance $L_J = \Phi_0 / (4\pi I_J )$). To verify that the broader coupling tunability of the LHST line discussed in Sec.~\ref{sec:frecsycoupls} is not an artifact of a specific parameter choice, Fig.~\ref{fig:barridoC_JyI_J} presents numerical sweeps over the $(C_J, I_J)$ parameter space while keeping the bulk parameters ($L$, $C$, and $N=30$) fixed.

The explored $(C_J, I_J)$ window covers realistic physical values reported in the experimental literature for superconducting SQUID architectures~\cite{apC1,apC2,apC3,apC4,apC5,apC6}. Evaluating the difference in maximum ($\Delta g_{\max} = g_{\max}^L - g_{\max}^R$) and minimum ($\Delta g_{\min} = g_{\min}^L - g_{\min}^R$) couplings confirms that $\Delta g_{\max} > 0$ and $\Delta g_{\min} < 0$ across the entire experimental regime. Thus, no realistic combination of boundary ratios $2C_J/C$ and $L/L_J$ inverts the spectral trend reported in the main text.

\section{Effective Hamiltonians and Rotating Frame Dynamics}
\label{app:rwa_and_rot}

In Secs. \ref{sec:singleresonance} and \ref{sec:multipleresonance} we outlined how the full parametrically driven Hamiltonian in Eq. \eqref{eq:Hleft} is reduced to the effective, time-independent models in Eqs. \eqref{eq:Honemode}, \eqref{eq:Htwomode}, and \eqref{eq:Hthreemode}. In this Appendix, we write out every intermediate step of this reduction---moving to the interaction picture, applying the Rotating Wave Approximation (RWA)~\cite{ScullyZubairy1997}, returning to the Schrödinger picture, and finally applying the rotating frame unitary transformations---for the two-mode resonance, and then list what changes for the other resonances. We also demonstrate that the expectation value of the modal number operator remains invariant under these transformations, justifying our computation of particle creation directly in the rotating frame.

We take as worked example the two-mode resonance (Eq. \eqref{eq:Htwomode}), where the pump frequency satisfies $\Omega \approx \omega_k + \omega_l$. 

Starting from the full Hamiltonian $\hat{H} = \hat{H}_0 + \hat{H}_{\rm int}(t)$ given in Eq. \eqref{eq:Hleft}, the unperturbed free-field Hamiltonian is $\hat{H}_0 = \sum_i \hbar\omega_i \hat{a}_i^\dagger \hat{a}_i$, and the time-dependent interaction, still in the Schrödinger picture, is

\begin{equation}
    \hat{H}_{\rm int}(t) = \hbar \sum_{i,j} g_{ij} \cos(\Omega t) (\hat{a}_i\hat{a}_j + \hat{a}_i\hat{a}_j^\dagger + \hat{a}_i^\dagger\hat{a}_j + \hat{a}_i^\dagger\hat{a}_j^\dagger).
\end{equation}
To isolate the resonant dynamics, we first move to the interaction picture with respect to $\hat{H}_0$, where the operators evolve as $\hat{a}_i(t) = \hat{a}_i e^{-i\omega_i t}$. Expanding the pump term as $\cos(\Omega t) = \frac{1}{2}(e^{i\Omega t} + e^{-i\Omega t})$, the interaction Hamiltonian becomes a sum of oscillatory terms. Under the RWA, we assume the coupling is weak compared to the bare frequencies ($g_{kl} \ll \omega_k, \omega_l$) and discard the terms that oscillate rapidly, at frequencies such as $\Omega + \omega_k + \omega_l$ or $\Omega$ itself, retaining only the slowly varying resonant terms. 

For the specific condition $\Omega \approx \omega_k + \omega_l$, the surviving terms are those corresponding to the simultaneous creation and annihilation of particle pairs in modes $k$ and $l$: 
\begin{equation}
    \hat{H}_{\rm int, RWA}^{(I)}(t) \approx \hbar g_{kl} \left( \hat{a}_k^\dagger \hat{a}_l^\dagger e^{-i(\Omega - \omega_k - \omega_l)t} + \hat{a}_k \hat{a}_l e^{i(\Omega - \omega_k - \omega_l)t} \right).
\end{equation}
Note that the factor of $1/2$ from the cosine is exactly canceled by the sum over the symmetric index pairs $(k,l)$ and $(l,k)$ in the original Hamiltonian, which give identical contributions since $g_{kl}=g_{lk}$ and operators of different modes commute.

We then map this RWA Hamiltonian back to the Schrödinger picture by reintroducing the free evolution $\hat{H}_0$. This yields
\begin{equation}
    \hat{H}_{\text{RWA}}^{(S)}(t) = \hbar\omega_k \hat{a}_k^\dagger \hat{a}_k + \hbar\omega_l \hat{a}_l^\dagger \hat{a}_l + \hbar g_{kl} \left( \hat{a}_k^\dagger \hat{a}_l^\dagger e^{-i\Omega t} + \hat{a}_k \hat{a}_l e^{i\Omega t} \right).
\end{equation}
To completely remove the explicit time dependence, we transform to a rotating frame defined by the unitary operator
\begin{equation}
    \hat{\mathcal{U}}(t) = \exp\left[-i \left( \omega_{\text{rot},k} \hat{a}_k^\dagger \hat{a}_k + \omega_{\text{rot},l} \hat{a}_l^\dagger \hat{a}_l \right) t \right].
\end{equation}
The effective Hamiltonian in this frame is given by $\hat{H}_{\text{eff}} = \hat{\mathcal{U}}^\dagger \hat{H}_{\text{RWA}}^{(S)} \hat{\mathcal{U}} - i\hbar \hat{\mathcal{U}}^\dagger \dot{\hat{\mathcal{U}}}$. Applying the operator transformations $\hat{\mathcal{U}}^\dagger \hat{a}_k \hat{\mathcal{U}} = \hat{a}_k e^{-i\omega_{\text{rot},k} t}$, we find:
\begin{align}
    \frac{\hat{H}_{\text{eff}}}{\hbar} = &\left(\omega_k - \omega_{\text{rot},k}\right)\hat{a}_k^\dagger \hat{a}_k + \left(\omega_l - \omega_{\text{rot},l}\right)\hat{a}_l^\dagger \hat{a}_l \nonumber \\
    &+ g_{kl} \left( \hat{a}_k^\dagger \hat{a}_l^\dagger e^{-i(\Omega - \omega_{\text{rot},k} - \omega_{\text{rot},l})t} + \text{h.c.} \right).
\end{align}
To make this Hamiltonian perfectly time-independent, we require $\omega_{\text{rot},k} + \omega_{\text{rot},l} = \Omega$. By defining the detuning $\Delta_{kl} = \Omega - (\omega_k + \omega_l)$ and symmetrically distributing it such that $\omega_{\text{rot},k} = \omega_k + \Delta_{kl}/2$ and $\omega_{\text{rot},l} = \omega_l + \Delta_{kl}/2$, we immediately recover Eq. \eqref{eq:Htwomode} of the main text:
\begin{equation}
    \frac{\hat{H}_{\text{eff}}}{\hbar} = -\frac{\Delta_{kl}}{2} \hat{a}_k^\dagger \hat{a}_k - \frac{\Delta_{kl}}{2} \hat{a}_l^\dagger \hat{a}_l + g_{kl} (\hat{a}_k^\dagger \hat{a}_l^\dagger + \hat{a}_k \hat{a}_l).
\end{equation}
The single-mode (Eq. \eqref{eq:Honemode}) and multi-mode (Eq. \eqref{eq:Hthreemode}) effective Hamiltonians are derived following identically the exact same steps, varying only the selection of resonant terms kept during the RWA and the choice of the respective rotating frequencies.

Finally, we address the computation of the mean particle number $\langle \hat{N}_i \rangle$ directly from the operators derived in this rotating frame. In the standard Heisenberg picture, the number operator evolves with the full system propagator. However, because our rotating frame unitary $\hat{\mathcal{U}}(t)$ is generated solely by a sum of number operators $\hat{N}_i = \hat{a}_i^\dagger \hat{a}_i$, it strictly commutes with $\hat{N}_i$ itself: $[\hat{\mathcal{U}}(t), \hat{N}_i] = 0$. 

Consequently, the number operator in the rotating frame remains identical to the laboratory frame number operator:
\begin{equation}
    \hat{N}_{\text{rot}, i} = \hat{\mathcal{U}}^\dagger(t) \hat{N}_i \hat{\mathcal{U}}(t) = \hat{N}_i.
\end{equation}
This implies that the expectation value computed in the rotating frame state $|\psi_{\text{rot}}(t)\rangle = \hat{\mathcal{U}}^\dagger(t) |\psi(t)\rangle$ exactly matches the physical laboratory observable:
\begin{align}
    \langle \hat{N}_i \rangle_{\text{rot}} &= \langle \psi_{\text{rot}}(t) | \hat{N}_i | \psi_{\text{rot}}(t) \rangle \nonumber \\
    &= \langle \psi(t) | \hat{\mathcal{U}}(t) \hat{N}_i \hat{\mathcal{U}}^\dagger(t) | \psi(t) \rangle \nonumber \\
    &= \langle \psi(t) | \hat{N}_i | \psi(t) \rangle = \langle \hat{N}_i \rangle_{\text{lab}}.
\end{align}
This exact structural invariance justifies bypassing the highly oscillatory inverse transformations back to the laboratory frame.  The same argument shows that the free choice of rotating frequencies in the three-mode case does not affect $\langle\hat N_i\rangle$: two admissible frames differ by a unitary generated by number operators. 

\section{Asymptotic Scaling of Frequency Spacing and Coupling Strength}
\label{app:asymptotic_scaling}

The purpose of this appendix is to analytically derive the asymptotic scaling of both the frequency spacing $\Delta\omega = \omega_i - \omega_{i+1}$ and the inter-mode coupling strength $g$ as the number of unit cells $N$ increases. We focus on the dense spectral region near the edge of the Brillouin zone.

\textit{Mode Distribution near the Brillouin Zone Edge:} Near the edge of the Brillouin zone, $\chi_i \rightarrow \pi$. To analyze a small deviation from this boundary, we introduce:
\begin{equation}
    \chi_i = \pi - \epsilon_i \quad \text{with} \quad \epsilon_i \ll 1 \text{.}
\end{equation}

For a transmission line terminated with specific boundary conditions, the allowed modes are governed by a transcendental equation. Expanding the boundary constraints for small $\epsilon_i$ to the first non-vanishing order yields an approximated equation of the form:
\begin{equation}
    \tan(N\epsilon_i) = -\frac{\epsilon_i}{\beta} \text{,}
    \label{eq:transcendental_approx}
\end{equation}
where $\beta$ is a constant determined by the circuit parameters. 

Since $\epsilon_i$ is small, the argument $N\epsilon_i$ must be close to a multiple of $\pi$. We can define an integer $m = N - i$ (representing the mode index relative to the band edge) and a small phase shift $\delta \ll 1$, such that:
\begin{equation}
    N\epsilon_i = m\pi - \delta \implies \epsilon_i = \frac{m\pi}{N} - \frac{\delta}{N} \text{.}
\end{equation}

Evaluating the tangent of this argument gives:
\begin{equation}
    \tan(m\pi - \delta) = -\tan(\delta) \approx -\delta \text{.}
\end{equation}
Equating this to the right-hand side of Eq. \eqref{eq:transcendental_approx}, we find:
\begin{equation}
    -\delta \approx -\frac{\epsilon_i}{\beta} \approx -\frac{m\pi}{N\beta} \implies \delta = \frac{m\pi}{N\beta} \text{.}
\end{equation}

Consequently, the wave vector deviation $\epsilon_i$ is explicitly given by:
\begin{equation}
    \epsilon_i = \frac{m\pi}{N} - \frac{m\pi}{N^2 \beta} = \frac{m\pi}{N}\left(1 - \frac{1}{N\beta}\right) \text{.}
\end{equation}
To leading order, we observe that the wave vector separation scales as $\epsilon_i \sim \mathcal{O}\left(\frac{1}{N}\right)$.

\textit{Asymptotic Scaling of $\Delta\omega$:} Near the boundary of the Brillouin zone, the dispersion relation is quadratic with respect to the deviation $\epsilon_i$. Thus, the frequency of mode $i$ near the minimum frequency $\omega_{\text{min}}$ can be approximated as:
\begin{equation}
    \omega_i \approx \omega_{\text{min}} \left( 1 + \frac{\epsilon_i^2}{8} \right) \approx \omega_{\text{min}} \left( 1 + \frac{m^2 \pi^2}{8 N^2} \right) \text{.}
\end{equation}
Evaluating this expression for the adjacent mode (changing from $i$ to $i+1$, which shifts the index $m$ to $m-1$), we get:
\begin{equation}
    \omega_{i+1} \approx \omega_{\text{min}} \left( 1 + \frac{(m-1)^2 \pi^2}{8 N^2} \right) \text{.}
\end{equation}
The frequency spacing between two adjacent modes near the band edge is therefore:
\begin{align}
    \Delta\omega = \omega_i - \omega_{i+1} \approx \frac{\omega_{\text{min}} \pi^2}{8 N^2} \left[ m^2 - (m-1)^2 \right] \nonumber\\= \frac{\omega_{\text{min}} \pi^2 (2m - 1)}{8 N^2} \text{.}
\end{align}
This demonstrates that the frequency gap scales as $\Delta\omega \sim \mathcal{O}\left(\frac{1}{N^2}\right)$ in the dense zone of the spectrum.

\textit{Asymptotic Scaling of the Coupling $g$:} The parametric coupling strength $g$ between two modes relies on their spatial overlap across the line and their respective normalization constants. 

First, the normalization for each mode requires summing over the $N$ unit cells, which contributes a factor of $\sqrt{N/2} \sim \mathcal{O}(\sqrt{N})$ to the denominator. When calculating the coupling between two modes, their combined normalization introduces a global scaling factor of roughly $1/N$.

Second, we evaluate the spatial overlap term in the numerator, which depends on the field values at the boundaries:
\begin{align}
    \sin(k_i N \Delta x) &= \sin(N\chi_i) = \sin(N\pi - N\epsilon_i) \nonumber \\
    &= (-1)^{N+1} \sin(N\epsilon_i) = (-1)^{N+1} \sin(m\pi - \delta) \nonumber \\
    &= (-1)^{N+1} (-1)^{m+1} \sin(\delta) \approx \pm \delta \approx \pm \frac{m\pi}{N\beta} \text{.}
\end{align}
For the cross-product of the spatial distributions of two interacting modes $i$ and $j$ (associated with indices $m_i$ and $m_j$), this yields:
\begin{equation}
    \sin(k_i N \Delta x) \sin(k_j N \Delta x) \approx \pm \frac{m_i m_j \pi^2}{N^2 \beta^2} \sim \mathcal{O}\left(\frac{1}{N^2}\right) \text{.}
\end{equation}

Finally, the cross-frequency term $\sqrt{\omega_i \omega_j}$ remains approximately constant at $\omega_{\text{min}}$. Combining all these contributions—the spatial overlap product scaling as $\sim 1/N^2$ and the joint normalization providing an additional $1/N$ factor—we obtain the final scaling behavior for the coupling strength:
\begin{equation}
    g \propto \frac{1}{N} \left( \frac{m_i m_j}{N^2 \beta^2} \right) = \frac{m_i m_j}{N^3 \beta^2} \sim \mathcal{O}\left(\frac{1}{N^3}\right) \text{.}
\end{equation}

This confirms that as $N \gg 1$, the coupling strength $g$ decays one power of $N$ faster than the frequency spacing $\Delta\omega$, preventing widespread multimode excitation in the dense regions of the spectrum.

\bibliography{references}

@article{FerreriWilhemCreation,
  title = {Particle creation in left-handed metamaterial transmission lines},
  author = {Ferreri, Alessandro and Bruschi, David Edward and Wilhelm, Frank K.},
  journal = {Phys. Rev. Res.},
  volume = {6},
  issue = {3},
  pages = {033204},
  numpages = {10},
  year = {2024},
  month = {Aug},
  publisher = {American Physical Society},
  doi = {10.1103/PhysRevResearch.6.033204},
  url = {https://link.aps.org/doi/10.1103/PhysRevResearch.6.033204}
}

@article{FerreriWilhemSimulation,
  title = {Quantum simulation in hybrid transmission lines},
  author = {Ferreri, Alessandro and Wilhelm, Frank K.},
  journal = {Phys. Rev. Appl.},
  volume = {23},
  issue = {2},
  pages = {024026},
  numpages = {16},
  year = {2025},
  month = {Feb},
  publisher = {American Physical Society},
  doi = {10.1103/PhysRevApplied.23.024026},
  url = {https://link.aps.org/doi/10.1103/PhysRevApplied.23.024026}
}

@article{EggerWilhem2013,
  title = {Multimode Circuit Quantum Electrodynamics with Hybrid Metamaterial Transmission Lines},
  author = {Egger, D. J. and Wilhelm, F. K.},
  journal = {Phys. Rev. Lett.},
  volume = {111},
  issue = {16},
  pages = {163601},
  numpages = {5},
  year = {2013},
  month = {Oct},
  publisher = {American Physical Society},
  doi = {10.1103/PhysRevLett.111.163601},
  url = {https://link.aps.org/doi/10.1103/PhysRevLett.111.163601}
}

@article{Messinger2019,
  title = {Left-handed superconducting metamaterials for quantum simulation},
  author = {Messinger, Andreas and Taketani, Bruno G. and Wilhelm, Frank K.},
  journal = {Phys. Rev. A},
  volume = {99},
  pages = {032325},
  year = {2019},
  doi = {10.1103/PhysRevA.99.032325}
}

@article{Wang2019,
  title = {Mode Structure in Superconducting Metamaterial Transmission-Line Resonators},
  author = {Wang, H. and Zhuravel, A. P. and Indrajeet, S. and Taketani, B. G. and Hutchings, M. D. and Hao, Y. and Rouxinol, F. and Wilhelm, F. K. and LaHaye, M. D. and Ustinov, A. V. and Plourde, B. L. T.},
  journal = {Phys. Rev. Appl.},
  volume = {11},
  pages = {054062},
  year = {2019},
  doi = {10.1103/PhysRevApplied.11.054062}
}

@article{Indrajeet2020,
  title = {Coupling a Superconducting Qubit to a Left-Handed Metamaterial Resonator},
  author = {Indrajeet, S. and Wang, H. and Hutchings, M. D. and Taketani, B. G. and Wilhelm, Frank K. and LaHaye, M. D. and Plourde, B. L. T.},
  journal = {Phys. Rev. Appl.},
  volume = {14},
  pages = {064033},
  year = {2020},
  doi = {10.1103/PhysRevApplied.14.064033}
}

@article{McBroomCarroll2024,
  title = {Entangling Interactions Between Artificial Atoms Mediated by a Multimode Left-Handed Superconducting Ring Resonator},
  author = {McBroom-Carroll, T. and Schlabes, A. and Xu, X. and Ku, J. and Cole, B. and Indrajeet, S. and LaHaye, M. D. and Ansari, M. H. and Plourde, B. L. T.},
  journal = {PRX Quantum},
  volume = {5},
  pages = {020325},
  year = {2024},
  doi = {10.1103/PRXQuantum.5.020325}
}

@ARTICLE{Eleftheriades2002,
  author={Eleftheriades, G.V. and Iyer, A.K. and Kremer, P.C.},
  journal={IEEE Transactions on Microwave Theory and Techniques}, 
  title={Planar negative refractive index media using periodically L-C loaded transmission lines}, 
  year={2002},
  volume={50},
  number={12},
  pages={2702-2712},
  doi={10.1109/TMTT.2002.805197}}

@article{Kozyrev2008,
  title = {Nonlinear left-handed transmission line metamaterials},
  author = {Kozyrev, Alexander B. and van der Weide, Daniel W.},
  journal = {J. Phys. D: Appl. Phys.},
  volume = {41},
  pages = {173001},
  year = {2008},
  doi = {10.1088/0022-3727/41/17/173001}
}

@article{LaiItohCaloz2004,
  title = {Composite right/left-handed transmission line metamaterials},
  author = {Lai, Anthony and Itoh, Tatsuo and Caloz, Christophe},
  journal = {IEEE Microwave Magazine},
  volume = {5},
  number = {3},
  pages = {34--50},
  year = {2004},
  doi = {10.1109/MMW.2004.1337766}
}

@article{Veselago1968,
	author = {V. G. Veselago},
	title = {The electrodynamics of substances with simultaneously negative values of $\varepsilon$ and $\mu$},
	year = {1968},
	journal = {Sov. Phys. Usp.},
	volume = {10},
	number = {4},
	pages = {509--514},
	url = {https://ufn.ru/en/articles/1968/4/d/},
	doi = {10.1070/PU1968v010n04ABEH003699}
}

@article{Smith2000,
  title = {Composite Medium with Simultaneously Negative Permeability and Permittivity},
  author = {Smith, D. R. and Padilla, Willie J. and Vier, D. C. and Nemat-Nasser, S. C. and Schultz, S.},
  journal = {Phys. Rev. Lett.},
  volume = {84},
  pages = {4184--4187},
  year = {2000},
  doi = {10.1103/PhysRevLett.84.4184}
}

@article{Pendry2000,
  title = {Negative Refraction Makes a Perfect Lens},
  author = {Pendry, J. B.},
  journal = {Phys. Rev. Lett.},
  volume = {85},
  issue = {18},
  pages = {3966--3969},
  year = {2000},
  month = {Oct},
  publisher = {American Physical Society},
  doi = {10.1103/PhysRevLett.85.3966},
  url = {https://link.aps.org/doi/10.1103/PhysRevLett.85.3966}
}

@article{Houck2003,
  title = {Experimental Observations of a Left-Handed Material That Obeys Snell's Law},
  author = {Houck, Andrew A. and Brock, Jeffrey B. and Chuang, Isaac L.},
  journal = {Phys. Rev. Lett.},
  volume = {90},
  pages = {137401},
  year = {2003},
  doi = {10.1103/PhysRevLett.90.137401}
}

@article{Caloz2004,
author = {Caloz, Christophe and Itoh, Tatsuo},
title = {Lossy transmission line metamaterials},
journal = {Microwave and Optical Technology Letters},
volume = {43},
number = {2},
pages = {112-114},
doi = {10.1002/mop.20392},
year = {2004}
}

@book{CalozItohBook,
  author = {C.~Caloz and T.~Itoh},
  title = {Electromagnetic Metamaterials: Transmission Line Theory and Microwave Applications},
  publisher = {John Wiley \& Sons},
  address = {Hoboken, NJ},
  isbn = {9780471754329},
  doi = {10.1002/0471754323},
  year = {2005}
}

@article{Casimir1948,
  author = {Casimir, H. B. G.},
  title = {On the attraction between two perfectly conducting plates},
  journal = {Proc. K. Ned. Akad. Wet.},
  volume = {51},
  pages = {793},
  year = {1948}
}

@article{Moore1970,
  author = {Moore, Gerald T.},
  title = {Quantum Theory of the Electromagnetic Field in a Variable-Length One-Dimensional Cavity},
  journal = {J. Math. Phys.},
  volume = {11},
  pages = {2679--2691},
  year = {1970},
  doi = {10.1063/1.1665432}
}

@article{Law1994,
  author = {Law, C. K.},
  title = {Resonance Response of the Quantum Vacuum to an Oscillating Boundary},
  journal = {Phys. Rev. Lett.},
  volume = {73},
  pages = {1931--1934},
  year = {1994},
  doi = {10.1103/PhysRevLett.73.1931}
}

@article{DodonovKlimov1996,
  author = {Dodonov, V. V. and Klimov, A. B.},
  title = {Generation and detection of photons in a cavity with a resonantly oscillating boundary},
  journal = {Phys. Rev. A},
  volume = {53},
  pages = {2664--2682},
  year = {1996},
  doi = {10.1103/PhysRevA.53.2664}
}

@article{Crocce2001,
  author = {Crocce, M. and Dalvit, D. A. R. and Mazzitelli, F. D.},
  title = {Resonant photon creation in a three-dimensional oscillating cavity},
  journal = {Phys. Rev. A},
  volume = {64},
  pages = {013808},
  year = {2001},
  doi = {10.1103/PhysRevA.64.013808}
}

@article{Crocce2002,
  author = {Crocce, M. and Dalvit, D. A. R. and Mazzitelli, F. D.},
  title = {Quantum electromagnetic field in a three-dimensional oscillating cavity},
  journal = {Phys. Rev. A},
  volume = {66},
  pages = {033811},
  year = {2002},
  doi = {10.1103/PhysRevA.66.033811}
}

@incollection{DalvitMaiaNetoMazzitelli2011,
  author = {Dalvit, D. A. R. and Maia Neto, P. A. and Mazzitelli, F. D.},
  title = {Fluctuations, Dissipation and the Dynamical Casimir Effect},
  booktitle = {Casimir Physics},
  series = {Lecture Notes in Physics},
  volume = {834},
  editor = {Dalvit, D. and Milonni, P. and Roberts, D. and da Rosa, F.},
  publisher = {Springer},
  address = {Berlin, Heidelberg},
  pages = {419--457},
  year = {2011},
  doi = {10.1007/978-3-642-20288-9_13}
}

@article{Nation2012,
  author = {Nation, P. D. and Johansson, J. R. and Blencowe, M. P. and Nori, Franco},
  title = {Colloquium: Stimulating uncertainty: Amplifying the quantum vacuum with superconducting circuits},
  journal = {Rev. Mod. Phys.},
  volume = {84},
  pages = {1--24},
  year = {2012},
  doi = {10.1103/RevModPhys.84.1}
}

@article{Dodonov2020,
  author = {Dodonov, Viktor},
  title = {Fifty Years of the Dynamical Casimir Effect},
  journal = {Physics},
  volume = {2},
  number = {1},
  pages = {67--104},
  year = {2020},
  doi = {10.3390/physics2010007}
}

@article{Johansson2009,
  author = {Johansson, J. R. and Johansson, G. and Wilson, C. M. and Nori, Franco},
  title = {Dynamical Casimir Effect in a Superconducting Coplanar Waveguide},
  journal = {Phys. Rev. Lett.},
  volume = {103},
  pages = {147003},
  year = {2009},
  doi = {10.1103/PhysRevLett.103.147003}
}

@article{JohanssonCasimir2010,
  title = {Dynamical Casimir effect in superconducting microwave circuits},
  author = {Johansson, J. R. and Johansson, G. and Wilson, C. M. and Nori, Franco},
  journal = {Phys. Rev. A},
  volume = {82},
  issue = {5},
  pages = {052509},
  numpages = {17},
  year = {2010},
  month = {Nov},
  publisher = {American Physical Society},
  doi = {10.1103/PhysRevA.82.052509},
  url = {https://link.aps.org/doi/10.1103/PhysRevA.82.052509}
}

@article{Wilson2010,
  author = {Wilson, C. M. and Duty, T. and Sandberg, M. and Persson, F. and Shumeiko, V. and Delsing, P.},
  title = {Photon Generation in an Electromagnetic Cavity with a Time-Dependent Boundary},
  journal = {Phys. Rev. Lett.},
  volume = {105},
  pages = {233907},
  year = {2010},
  doi = {10.1103/PhysRevLett.105.233907}
}

@article{Wilson2011,
  author = {Wilson, C. M. and Johansson, G. and Pourkabirian, A. and Simoen, M. and Johansson, J. R. and Duty, T. and Nori, F. and Delsing, P.},
  title = {Observation of the dynamical Casimir effect in a superconducting circuit},
  journal = {Nature},
  volume = {479},
  pages = {376--379},
  year = {2011},
  doi = {10.1038/nature10561}
}

@article{Lahteenmaki2013,
  author = {L{\"a}hteenm{\"a}ki, Pasi and Paraoanu, G. S. and Hassel, Juha and Hakonen, Pertti J.},
  title = {Dynamical Casimir effect in a Josephson metamaterial},
  journal = {Proc. Natl. Acad. Sci. USA},
  volume = {110},
  pages = {4234--4238},
  year = {2013},
  doi = {10.1073/pnas.1212705110}
}

@article{Svensson2018,
  author = {Svensson, Ida-Maria and Bengtsson, Andreas and Bylander, Jonas and Shumeiko, Vitaly and Delsing, Per},
  title = {Period multiplication in a parametrically driven superconducting resonator},
  journal = {Appl. Phys. Lett.},
  volume = {113},
  pages = {022602},
  year = {2018},
  doi = {10.1063/1.5026974}
}

@article{Schneider2020,
  author = {Schneider, B. H. and Bengtsson, A. and Svensson, I. M. and Aref, T. and Johansson, G. and Bylander, J. and Delsing, P.},
  title = {Observation of Broadband Entanglement in Microwave Radiation from a Single Time-Varying Boundary Condition},
  journal = {Phys. Rev. Lett.},
  volume = {124},
  pages = {140503},
  year = {2020},
  doi = {10.1103/PhysRevLett.124.140503}
}

@article{Wustmann2013,
  author = {Wustmann, Waltraut and Shumeiko, Vitaly},
  title = {Parametric resonance in tunable superconducting cavities},
  journal = {Phys. Rev. B},
  volume = {87},
  pages = {184501},
  year = {2013},
  doi = {10.1103/PhysRevB.87.184501}
}

@article{CastellanosBeltran2008,
  author = {Castellanos-Beltran, M. A. and Irwin, K. D. and Hilton, G. C. and Vale, L. R. and Lehnert, K. W.},
  title = {Amplification and squeezing of quantum noise with a tunable Josephson metamaterial},
  journal = {Nat. Phys.},
  volume = {4},
  pages = {929--931},
  year = {2008},
  doi = {10.1038/nphys1090}
}

@article{Macklin2015,
  author = {Macklin, C. and O'Brien, K. and Hover, D. and Schwartz, M. E. and Bolkhovsky, V. and Zhang, X. and Oliver, W. D. and Siddiqi, I.},
  title = {A near-quantum-limited Josephson traveling-wave parametric amplifier},
  journal = {Science},
  volume = {350},
  pages = {307--310},
  year = {2015},
  doi = {10.1126/science.aaa8525}
}

@article{Lombardo2016,
  author = {Lombardo, F. C. and Mazzitelli, F. D. and Soba, A. and Villar, P. I.},
  title = {Dynamical Casimir effect in superconducting circuits: A numerical approach},
  journal = {Phys. Rev. A},
  volume = {93},
  pages = {032501},
  year = {2016},
  doi = {10.1103/PhysRevA.93.032501}
}

@article{Lombardo2018,
  author = {Lombardo, F. C. and Mazzitelli, F. D. and Soba, A. and Villar, P. I.},
  title = {Dynamical Casimir effect in a double tunable superconducting circuit},
  journal = {Phys. Rev. A},
  volume = {98},
  pages = {022512},
  year = {2018},
  doi = {10.1103/PhysRevA.98.022512}
}

@article{DelGrosso2020,
  author = {Del Grosso, N. F. and Lombardo, F. C. and Villar, P. I.},
  title = {Entanglement degradation of cavity modes due to the dynamical Casimir effect},
  journal = {Phys. Rev. D},
  volume = {102},
  pages = {125008},
  year = {2020},
  doi = {10.1103/PhysRevD.102.125008}
}

@article{DelGrosso2022Otto,
  author = {Del Grosso, N. F. and Lombardo, F. C. and Mazzitelli, F. D. and Villar, P. I.},
  title = {Quantum Otto cycle in a superconducting cavity in the nonadiabatic regime},
  journal = {Phys. Rev. A},
  volume = {105},
  pages = {022202},
  year = {2022},
  doi = {10.1103/PhysRevA.105.022202}
}

@article{Velasco2022,
  author = {Velasco, C. I. and Del Grosso, N. F. and Lombardo, F. C. and Soba, A. and Villar, P. I.},
  title = {Photon generation and entanglement in a double superconducting cavity},
  journal = {Phys. Rev. A},
  volume = {106},
  pages = {043701},
  year = {2022},
  doi = {10.1103/PhysRevA.106.043701}
}

@article{LouysSanso2025,
  author = {Louys Sans{\'o}, J. P. and Del Grosso, N. F. and Lombardo, F. C. and Villar, P. I.},
  title = {Superconducting quantum circuit to simulate the dynamical Casimir effect in a double cavity},
  journal = {Phys. Rev. A},
  volume = {111},
  pages = {013714},
  year = {2025},
  doi = {10.1103/PhysRevA.111.013714}
}

@article{DelGrosso2025Gate,
  author = {Del Grosso, N. F. and Corti{\~n}as, R. G. and Villar, P. I. and Lombardo, F. C. and Paz, J. P.},
  title = {Controlled-squeeze gate in superconducting quantum circuits},
  journal = {Phys. Rev. A},
  volume = {111},
  pages = {042606},
  year = {2025},
  doi = {10.1103/PhysRevA.111.042606}
}

@misc{LombardoVillar2025,
  author = {Lombardo, Fernando C. and Villar, Paula I.},
  title = {Dynamical Casimir effect in superconducting cavities: from photon generation to universal quantum gates},
  year = {2025},
  eprint = {2504.11361},
  archivePrefix = {arXiv},
  primaryClass = {quant-ph}
}

@article{YurkeDenker1984,
  author = {Yurke, B. and Denker, J. S.},
  title = {Quantum network theory},
  journal = {Phys. Rev. A},
  volume = {29},
  pages = {1419--1437},
  year = {1984},
  doi = {10.1103/PhysRevA.29.1419}
}

@article{VoolDevoret2017,
  author = {Vool, Uri and Devoret, Michel},
  title = {Introduction to quantum electromagnetic circuits},
  journal = {Int. J. Circuit Theory Appl.},
  volume = {45},
  pages = {897--934},
  year = {2017},
  doi = {10.1002/cta.2359}
}

@article{Blais2004,
  author = {Blais, Alexandre and Huang, Ren-Shou and Wallraff, Andreas and Girvin, S. M. and Schoelkopf, R. J.},
  title = {Cavity quantum electrodynamics for superconducting electrical circuits: An architecture for quantum computation},
  journal = {Phys. Rev. A},
  volume = {69},
  pages = {062320},
  year = {2004},
  doi = {10.1103/PhysRevA.69.062320}
}

@article{Wallraff2004,
  author = {Wallraff, A. and Schuster, D. I. and Blais, A. and Frunzio, L. and Huang, R.-S. and Majer, J. and Kumar, S. and Girvin, S. M. and Schoelkopf, R. J.},
  title = {Strong coupling of a single photon to a superconducting qubit using circuit quantum electrodynamics},
  journal = {Nature},
  volume = {431},
  pages = {162--167},
  year = {2004},
  doi = {10.1038/nature02851}
}

@article{Blais2021,
  author = {Blais, Alexandre and Grimsmo, Arne L. and Girvin, S. M. and Wallraff, Andreas},
  title = {Circuit quantum electrodynamics},
  journal = {Rev. Mod. Phys.},
  volume = {93},
  pages = {025005},
  year = {2021},
  doi = {10.1103/RevModPhys.93.025005}
}

@article{Wallquist2006,
  author = {Wallquist, M. and Shumeiko, V. S. and Wendin, G.},
  title = {Selective coupling of superconducting charge qubits mediated by a tunable stripline cavity},
  journal = {Phys. Rev. B},
  volume = {74},
  pages = {224506},
  year = {2006},
  doi = {10.1103/PhysRevB.74.224506}
}

@article{Weissl2015,
  author = {Wei{\ss}l, T. and K{\"u}ng, B. and Dumur, E. and Feofanov, A. K. and Matei, I. and Naud, C. and Buisson, O. and Hekking, F. W. J. and Guichard, W.},
  title = {Kerr coefficients of plasma resonances in Josephson junction chains},
  journal = {Phys. Rev. B},
  volume = {92},
  pages = {104508},
  year = {2015},
  doi = {10.1103/PhysRevB.92.104508}
}

@book{ScullyZubairy1997,
  author = {Scully, Marlan O. and Zubairy, M. Suhail},
  title = {Quantum Optics},
  publisher = {Cambridge University Press},
  address = {Cambridge},
  year = {1997},
  doi = {10.1017/CBO9780511813993}
}

@article{apC1,
  author  = {Tesche, Claudia D. and Clarke, John},
  title   = {dc {SQUID}: Noise and optimization},
  journal = {Journal of Low Temperature Physics},
  year    = {1977},
  volume  = {29},
  number  = {3},
  pages   = {301--331},
  doi     = {10.1007/BF00655097},
  issn    = {1573-7357},
  url     = {https://doi.org/10.1007/BF00655097}
}

@book{apC2,
  author  = {J. Clarke and A. I. Braginski},
  publisher = {Wiley-VCH},
  isbn = {9783527603640},
  title = {The SQUID Handbook},
  doi = {10.1002/3527603646},
  year = {2004}
}

@article{apC3,
    author = {Deppe, Frank and Saito, Shiro and Tanaka, Hirotaka and Takayanagi, Hideaki},
    title = {Determination of the capacitance of nm scale Josephson junctions},
    journal = {Journal of Applied Physics},
    volume = {95},
    number = {5},
    pages = {2607-2613},
    year = {2004},
    month = {03},
    issn = {0021-8979},
    doi = {10.1063/1.1645673},
    url = {https://doi.org/10.1063/1.1645673}
}

@article{apC4,
title = {Sub-fT/Hz1/2 resolution and field-stable SQUID magnetometer based on low parasitic capacitance sub-micrometer cross-type Josephson tunnel junctions},
journal = {Physica C: Superconductivity and its Applications},
volume = {482},
pages = {27-32},
year = {2012},
issn = {0921-4534},
doi = {10.1016/j.physc.2012.06.005},
author = {M. Schmelz and R. Stolz and V. Zakosarenko and T. Sch{\"o}nau and S. Anders and L. Fritzsch and M. M{\"u}ck and M. Meyer and H.-G. Meyer}
}

@article{apC5,
  author  = {Faley, M. I. and Kostyurina, E. A. and Kalashnikov, K. V. and Maslennikov, Yu. V. and Koshelets, V. P. and Dunin-Borkowski, R. E.},
  title   = {Superconducting Quantum Interferometers for Nondestructive Evaluation},
  journal = {Sensors},
  year    = {2017},
  volume  = {17},
  number  = {12},
  pages   = {2798},
  month   = dec,
  doi     = {10.3390/s17122798},
  issn    = {1424-8220},
  url     = {https://doi.org/10.3390/s17122798}
}

@article{apC6,
  author  = {Zheng, Yuzhen and Li, Shuming and Ding, Zengqian and Xiong, Kanglin and Feng, Jiagui and Yang, Hui},
  title   = {Fabrication of Al/AlOx/Al junctions with high uniformity and stability on sapphire substrates},
  journal = {Scientific Reports},
  year    = {2023},
  volume  = {13},
  number  = {1},
  pages   = {11874},
  doi     = {10.1038/s41598-023-39052-2},
  issn    = {2045-2322},
  url     = {https://doi.org/10.1038/s41598-023-39052-2}
}

@article{Urtubey2026,
  title = {Dissipative effects in transmission line analogs of Hawking radiation},
  author = {Urtubey, Eli{\'a}n and Lombardo, Fernando C. and Villar, Paula I.},
  journal = {Phys. Rev. D},
  volume = {114},
  issue = {8},
  pages = {085012},
  numpages = {12},
  year = {2026},
  month = {Oct},
  publisher = {American Physical Society},
  doi = {10.1103/g8b1-znb1},
  url = {https://link.aps.org/doi/10.1103/g8b1-znb1}
}

\end{document}